\documentclass[a4paper,twocolumn,11pt,accepted=2026-01-20]{quantumarticle}
\pdfoutput=1
\usepackage[utf8]{inputenc}
\usepackage[english]{babel}
\usepackage[T1]{fontenc}
\usepackage{physics}
\usepackage{amsmath}
\usepackage{amsfonts}
\usepackage{amssymb}
\usepackage{dsfont}
\usepackage{hhline}
\usepackage[abs]{overpic}
\usepackage{hyperref}
\usepackage[numbers]{natbib}

\usepackage{tikz}
\usepackage{lipsum}

\begin{document}

\title{Multi-qubit Rydberg gates between distant atoms}

\author{Antonis Delakouras}
\affiliation{Institute of Electronic Structure and Laser and Center for Quantum Science and Technologies, FORTH, 70013 Heraklion, Crete, Greece}
\affiliation{Department of Physics, University of Crete,
  Heraklion, Greece}

\author{Georgios Doultsinos}
\affiliation{Institute of Electronic Structure and Laser and Center for Quantum Science and Technologies, FORTH, 70013 Heraklion, Crete, Greece}
\affiliation{Department of Physics, University of Crete,
Heraklion, Greece}

\author{David Petrosyan}
\affiliation{Institute of Electronic Structure and Laser and Center for Quantum Science and Technologies, FORTH, 70013 Heraklion, Crete, Greece}
\affiliation{A. Alikhanyan National Science Laboratory (YerPhI), 0036 Yerevan, Armenia}
\orcid{0000-0002-9256-1096}

\maketitle

\begin{abstract}
We propose an efficient protocol to realize multi-qubit gates in arrays of neutral atoms. 
The atoms encode qubits in the long-lived hyperfine sublevels of the ground electronic state. 
To realize the gate, we apply a global laser pulse to transfer the atoms to a Rydberg state with strong blockade interaction that suppresses simultaneous excitation of neighboring atoms arranged in a star-graph configuration. 
The number of Rydberg excitations, and thereby the parity of the resulting state, depends on the multiqubit input state. 
Upon changing the sign of the interaction and de-exciting the atoms with an identical laser pulse, the system acquires a geometric phase that depends only on the parity of the excited state, while the dynamical phase is completely canceled. 
Using single qubit rotations, this transformation can be converted to the C$_k$Z or C$_k$NOT quantum gate for $k+1$ atoms. 
We also present extensions of the scheme to implement quantum gates between distant atomic qubits connected by a quantum bus consisting of a chain of atoms. 
\end{abstract}

\section{Introduction} 
\label{Sec:Intro}

Neutral atoms in arrays of reconfigurable mictotraps \cite{barredo2016atom,endres2016atom,barredo2018synthetic} excited by lasers to strongly interacting Rydberg states represent a highly versatile system for analog quantum simulations of many-body physics \cite{browaeys2020many,morgado2021quantum,labuhn2016tunable, Bernien2017, Keesling2019, Bakr2018, Lienhard2018, Scholl2021, ebadi2021quantum, tzortzakakis2022microscopic, kim2024realization, shaw2024, gonzalez2025observation} and digital quantum information processing and quantum computing \cite{Saffman2010InformationRydberg, xia2015, Levine2019Parallel, graham2019, graham2022, evered2023high, ma2023high, Tsai2025, MunizPRXQ.2025}.
Regular arrays of atoms subject to amplitude and frequency tunable lasers coupling the ground and Rydberg states can emulate the quantum Ising model for spins in effective longitudinal and transverse magnetic fields \cite{labuhn2016tunable, Bernien2017, Keesling2019, Bakr2018, Lienhard2018, Scholl2021, ebadi2021quantum, tzortzakakis2022microscopic}, enabling simulations and studies of quantum phases and phase transitions of interacting many-body systems.
Using the Rydberg excitation blockade \cite{jaksch2000fast,gaetan2009,Urban2009ObservationBlockade,WilkPRL2010RGate,IsenhowerPRL2010CNOTgate}, atoms arranged in various spatial configurations are used to solve hard optimization problems that can be mapped onto the maximal independent set (MIS) solutions for graphs of different connectivity \cite{ebadi2022quantum,AhnPhysRevResearch2021,kim2022rydberg}.
Meanwhile, the long lifetimes of Rydberg states and their strong interactions enable realizations of high-fidelity quantum gates in neutral atom quantum computers \cite{evered2023high, ma2023high, Tsai2025, MunizPRXQ.2025}. 

The dipole-dipole (DD) or van der Waals (vdW) interactions between the atoms in Rydberg states permit the realization of gates between pairs of atoms separated by less than the Rydberg blockade distance of several micrometers \cite{gaetan2009,Urban2009ObservationBlockade}. 
The gates are performed either by individual atom addressing \cite{Saffman2010InformationRydberg, jaksch2000fast, WilkPRL2010RGate, IsenhowerPRL2010CNOTgate}, or using laser pulses illuminating pairs of atoms simultaneously \cite{Saffman2020SymmetricCZ,Levine2019Parallel,jandura2022time,ma2023high, Tsai2025, MunizPRXQ.2025}. 
The long-range interactions between the Rydberg-excited atoms also enable implementations of native multiqubit gates \cite{KhazaliPRX2020, Levine2019Parallel, jandura2022time, cao2402multi, evered2023high, pelegri2022high, Petrosyan2024FastMultiqubit} that can significantly reduce the number of two-qubit gates required for quantum algorithms \cite{martinez2016compiling,KMJPhysB2011,DPJPhysB2016} and increase the circuit depth. 
Multiqubit gates are also highly advantageous for logical qubit encoding/decoding and error syndrome detection and correction in fault-tolerant quantum computation \cite{QCQI2000, xu2024constant, bluvstein2024logical, Reichardt2024}.   
To implement quantum gates between distant atoms, they moved during the computation to be brought within the interaction range \cite{bluvstein2022quantum}, which should, however, be done sufficiently slowly to avoid atom loss or motional decoherence. 

%%%%%%%%%%%%%%%%%%%%%%%%%%%%%
\begin{figure}[t] 
    \includegraphics[width=1\columnwidth]{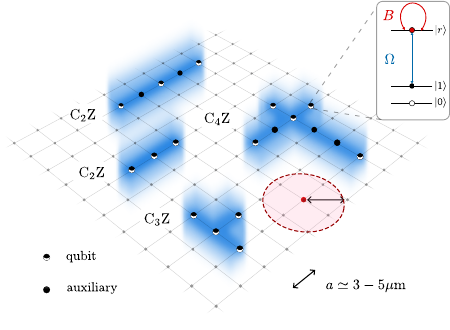} 
    \caption{Schematic illustration of a two-dimensional array of atoms irradiated by lasers (shaded blue) to implement the C$_k$Z quantum gates between neighboring atomic qubits in star-graph configurations, or between distant atomic qubits connected by auxiliary atoms prepared initially in state $\ket{1}$. 
    Red transparent circle around a Rydberg excited atom corresponds to blockade range.
    Inset shows the level scheme of atoms involving the qubit encoding ground state subleveles $\ket{0},\ket{1}$ and the strongly interacting Rydberg state $\ket{r}$. }
    \label{Fig:Intro}
\end{figure}
%%%%%%%%%%%%%%%%%%%%%%%%%%%%%

It is thus important to develop robust schemes to realize fast and reliable multiqubit quantum gates between distant atoms to achieve high connectivity in neutral atom quantum processors, which is the objective of the present work. 
We propose a novel scheme to implement multiqubit C$_k$Z gates between $k+1$ atoms driven by global laser pulses, as illustrated schematically in Fig.~\ref{Fig:Intro}.
In our protocol, as in most other schemes \cite{xia2015, Levine2019Parallel, graham2019, graham2022, evered2023high, ma2023high, Tsai2025, ma2023high}, the atoms encode qubits in a pair of long-lived hyperfine ground state sublevels $\ket{0},\ket{1}$. 
The atoms participating in a gate are positioned in star-graph configuration with strong Rydberg state interaction between the central atom and $k$ outer atoms, and weak interaction between the outer atoms. 
We apply a global laser pulse uniformly irradiating the participating atoms and coupling their qubit state $\ket{1}$ to the Rydberg state $\ket{r}$, while the other qubit state $\ket{0}$ remains passive. 
Starting with an input configuration $\ket{\mathbf{q}}= \ket{q_0 q_1 \ldots q_k}$ ($q_i=\{0,1\}$) of the states of all the qubits, upon adiabatic sweep of the laser detuning from the red to the blue side of the ground-Rydberg resonance, the atoms in the active qubit state are transferred to the state $\ket{R_{\mathbf{q}}}$ corresponding to the MIS solution for the given subgraph and containing a certain number $\nu_{\mathbf{q}}$ of Rydberg excitations that depends on the input $\ket{\mathbf{q}}$. 
Switching the sign of the interaction between the atoms in the Rydberg state and applying a reverse transformation with an identical chirped laser pulse, the atoms return to their initial state $(-1)^{\nu_{\mathbf{q}}}\ket{\mathbf{q}}$, but the combined state of the system acquires a geometric phase that depends only on the parity of $\nu_{\mathbf{q}}$ while the dynamical phase cancels exactly.
We show that, upon applying single-qubit X and Z gates to all but one qubit, this transformation is equivalent to the C$_k$Z gate. 
We also propose extensions of the scheme to implement long-range quantum gates between distant atomic qubits connected to the central qubit by a quantum bus realized by a chain of atoms initially in state $\ket{1}$ \cite{Doultsinos2025PRR}.

The paper is organized as follows.
In Sec.~\ref{Sec:MBS} we introduce the physical model for laser-driven atoms in a lattice, the principles of operation of multiqubit gates, and the dynamics of the many-body system.  
In Sec.~\ref{Sec:Fidelity} we analyze the resulting gate fidelity, taking into account the decay and non-adiabatic leakage errors, as well as errors due to Rydberg state transfer and thermal motion, and present pulse optimization for faster and better gates. 
In Sec. \ref{Sec:Distant_atoms} we discuss implementations of quantum gates between distant atomic qubits. 
Our conclusions are summarized in Sec.~\ref{Sec:Conclude}.

\section{The many-body system}
\label{Sec:MBS}

\subsection{Model} \label{Sec:Model}

We consider $N$ neutral atoms trapped by optical tweezers at positions $\mathbf{x}_i$ ($i=0,1,2, \ldots, N-1$) on a 2D plane.
Each atom has three relevant internal states to represent a qubit and interact with the other atoms. 
The qubit basis states $\ket{0}$ and $\ket{1}$ are encoded in a pair of long-lived hyperfine sub-levels of the electronic ground state, while a highly excited Rydberg state $\ket{r}$ mediates interactions between the atoms. 
A global laser field couples the state $\ket{1}$ of each atom to the Rydberg state $\ket{r}$ with time-dependent Rabi frequency $\Omega(t)$ detuning $\Delta(t) = \omega - \omega_{r1}$, while atoms in state $\ket{0}$ remain decoupled from the laser. 
In the frame rotating with the laser frequency $\omega$, the Hamiltonian of the system is ($\hbar=1$) 
\begin{align} \label{Eq:Hamiltonian}
  \mathcal{H}(t) = &- \Delta(t)\sum_{i=0}^{N-1}\ketbra{r_i}{r_i} \nonumber\\
  &+ \tfrac{1}{2}\Omega(t)\sum_{i=0}^{N-1} \left(\ketbra{r_i}{1_i}+\text{H.c.}\right) \nonumber\\
  &+\sum_{i<j} B_{ij}\ketbra{r_ir_j}{r_ir_j} ,
\end{align}
where $B_{ij}= \frac{C_6}{|\mathbf{x}_i-\mathbf{x}_j|^6}$ is the strength of the van der Waals interaction between the Rydberg-state atoms at positions $\mathbf{x}_i$ and $\mathbf{x}_j$ with $C_6$ being the interaction coefficient that depends on the selected Rydberg state $\ket{r}$. 
Interactions between the atoms in ground-state sublevels $\ket{0,1}$ are vanishingly small and can be neglected.

%%%%%%%%%%%%%%%%%%%%%%%%%%%%%
\begin{figure}[t] 
    \centering
    \includegraphics[width=0.9\columnwidth]{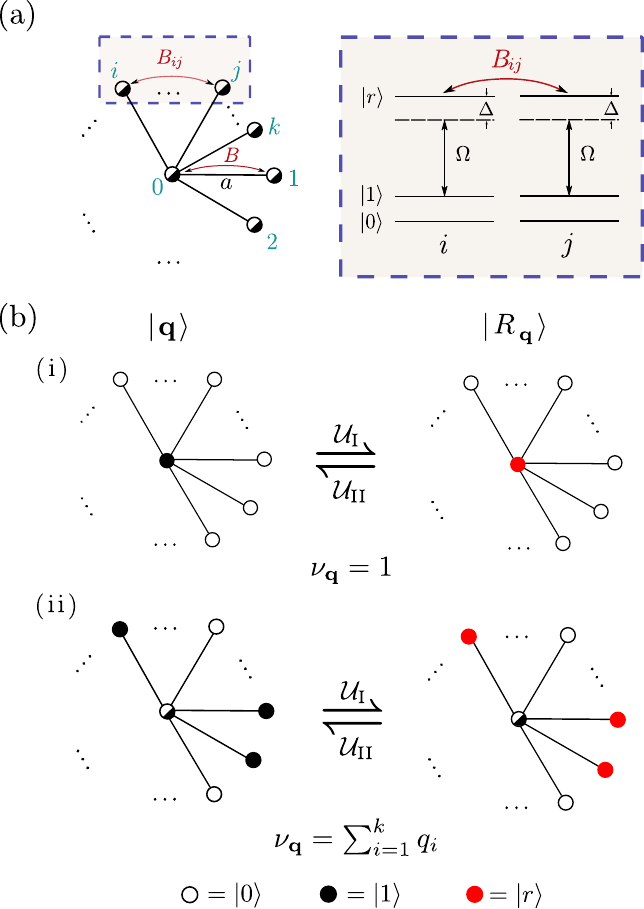} 
    \caption{(a) Star-graph configuration of $N=k+1$ atoms. 
    Half-filled circles represent atomic qubits in either state $\ket{0}$ or $\ket{1}$. 
    A global laser field with Rabi frequency $\Omega$ and detuning $\Delta$ couples the atomic states $\ket{1}$ and $\ket{r}$ while atoms in state $\ket{0}$ are decoupled from the laser (inset).
    Atoms in Rydberg state $\ket{r}$ interact with each other. Graph edges denote Rydberg blockade interactions between the connected atoms, $B= B_{0j}\gg\max(\Delta,\Omega)$, while absence thereof implies weak-interactions, $B_{ij} \ll\max(\Omega)$ for $i,j>0$. 
    (b) Schematic representation of the computational basis states of the atoms $\ket{\mathbf{q}}$, which is the ground state of $\mathcal{H}$ for $-\Delta \gg \Omega$, whereas the corresponding MIS configurations $\ket{R_{\mathbf{q}}}$ is the ground state for $\Delta \gg \Omega$. }
    \label{Fig:Model}
\end{figure}
%%%%%%%%%%%%%%%%%%%%%%%%%%%%%

We consider star-graph configurations of $N=k+1$ atoms, see Fig. \ref{Fig:Model}(a), assuming strong Rydberg-state interactions between the central atom and $k$ outer atoms, $B= B_{0j}\gg\max(\Delta,\Omega)$, whereas the interactions between the outer atoms are weak, $B_{ij} \ll\max(\Omega)$ for $i,j \geq 1$. 
Thus, simultaneous laser excitation of strongly interacting atoms is suppressed, while weakly interacting atoms can simultaneously be excited to the Rydberg state. 
Atoms in state $\ket{0}$ remain decoupled from the rest of the graph. 

Under such blockade interactions, the ground state of the Hamiltonian for large negative detuning $-\Delta\gg\max(\Omega)$ contains no atom in the Rydberg state and is of the form $\ket{\mathbf{q}}=\ket{q_0q_1...q_k}$, with $q_i=\{0,1\}$. 
Hence, every input of the computational basis corresponds to the ground state of a specific sub-graph for $\Delta<0$.
Conversely, for large positive detuning, $\Delta\gg\max(\Omega)$, the ground state of the system $\ket{R_{\mathbf{q}}}$ corresponding to input $\ket{\mathbf{q}}$ contains the maximum number $\nu_{\mathbf{q}}$ of Rydberg excitations of disconnected atoms, i.e., it is the MIS solution of the corresponding sub-graph. 
The number of Rydberg excitations in state $\ket{R_{\mathbf{q}}}$ is given by
\begin{equation} \label{Eq:nu_r-star}
    \nu_{\mathbf{q}} =
    \begin{cases} 
        1 & \text{if } q_0=1 \; \& \; q_i=0 \; \forall \;  i \in [1,k]  \\
        \sum_{i=1}^k q_i & \text{otherwise}, 
    \end{cases}
\end{equation}
as illustrated in Fig.~\ref{Fig:Model}(b): 
(i) If only the central atom $i=0$ is initially in state $\ket{1}$ and all the other atoms $i=1,2,\ldots,k$ are in the passive state $\ket{0}$, then the central atom can be excited to the Rydberg state $\ket{r}$, leading to $\nu_{\mathbf{q}}=1$; 
(ii) If one or more outer atoms are in state $\ket{1}$, then they can all be excited to the Rydberg state $\ket{r}$ while the central atom remains in its original state, resulting in $\nu_{\mathbf{q}} = \sum_{i=1}^k q_i$.

\subsection{The gate protocol} \label{Sec:Gate-protocol}

The multiqubit C$_k$Z gate corresponds to the transformation 
\begin{equation} \label{Eq:C_kZ-truth-table}
    \ket{q_0q_1...q_k}\to (-1)^{\prod^{k}_{i=0} q_i}\ket{q_0q_1...q_k}, \;\; q_i=\{0,1\} ,
\end{equation}
where the uniform input $\ket{11...1}$ acquires a sign flip, whereas all the other configurations remain unchanged. 

%%%%%%%%%%%%%%%%%%%%%%%%%%%%%%%%%%%%%%%%%%%%%%5
\begin{table}[t]
\caption{Illustration of the transformations $\mathcal{U}_{\mathrm{II}}\mathcal{U}_{\mathrm{I}}\ket{\mathbf{q}}$ of Eq.~(\ref{Eq:Pulse_Transformation}) for $N=3$ atoms and the resulting parity $(-1)^{\nu_{\mathbf{q}}}$.}
    \begin{tabular}{c c c c c }
        \hhline{=====}
        $\ket{q_0 q_1 q_2}$ & $\ket{\mathbf{q}}$ & $\rightleftharpoons$ & $\ket{R_{\mathbf{q}}}$ & Parity \\ 
        \hline
        $\ket{000}$ & $\circ$-$\circ$-$\circ$ & & $\circ$-$\circ$-$\circ$ & $+1$ \\
        $\ket{100}$ & $\circ$-$\bullet$-$\circ$ &  & $\circ$-$\textcolor{red}{\bullet}$-$\circ$ & $-1$ \\
        $\ket{010}$ & $\circ$-$\circ$-$\bullet$ &  & $\circ$-$\circ$-$\textcolor{red}{\bullet}$ & $-1$ \\
        $\ket{001}$ & $\bullet$-$\circ$-$\circ$ &  & $\textcolor{red}{\bullet}$-$\circ$-$\circ$ & $-1$\\  
        $\ket{110}$ & $\circ$-$\bullet$-$\bullet$ &  & $\frac{1}{\sqrt{2}}(\circ$-$\bullet$-$\textcolor{red}{\bullet}$ + $\circ$-$\textcolor{red}{\bullet}$-$\bullet$) & $-1$ \\
        $\ket{101}$ & $\bullet$-$\bullet$-$\circ$ &  & $\frac{1}{\sqrt{2}}(\bullet$-$\textcolor{red}{\bullet}$-$\circ$ + $\textcolor{red}{\bullet}$-$\bullet$-$\circ$) & $-1$ \\
        $\ket{011}$ & $\bullet$-$\circ$-$\bullet$ & & $\textcolor{red}{\bullet}$-$\circ$-$\textcolor{red}{\bullet}$ & $+1$ \\ 
        $\ket{111}$ & $\bullet$-$\bullet$-$\bullet$ & & $\textcolor{red}{\bullet}$-$\bullet$-$\textcolor{red}{\bullet}$ & $+1$ \\
        \hhline{=====}
    \end{tabular}
    \label{Tab:C2Z}
\end{table}
%%%%%%%%%%%%%%%%%%%%%%%%%%%%%%%%%%%%%%%%%%%%%%%%%

We define the parity operator $\mathcal{P}=\prod_i(\ketbra{1_i}{1_i}-\ketbra{r_i}{r_i})$ acting as 
\begin{subequations} \label{Eq:Parity}
    \begin{align}
        \mathcal{P}\ket{\mathbf{q}} &= \ket{\mathbf{q}}, \\
        \mathcal{P}\ket{R_{\mathbf{q}}}&=(-1)^{\nu_{\mathbf{q}}}\ket{R_{\mathbf{q}}}. 
    \end{align}
\end{subequations}
Our protocol to realize the gate relies on the sequence of global laser pulses that implement unitary transformations $\mathcal{U}_{\mathrm{I}}$ and $\mathcal{U}_{\mathrm{II}}=\mathcal{P}\mathcal{U}_{\mathrm{I}}^{-1}\mathcal{P}$ with the result 
\begin{equation}  \label{Eq:Pulse_Transformation}
    \ket{\mathbf{q}} \overset{\mathcal{U}_{\mathrm{I}}}{\longrightarrow}\ket{R_{\mathbf{q}}}\overset{\mathcal{U}_{\mathrm{II}}}{\longrightarrow} e^{i\phi_g}\ket{\mathbf{q}}, 
\end{equation}
where $\phi_g$ is a geometric phase resulting from the parity of state $\ket{R_{\mathbf{q}}}$, $e^{i\phi_g} = (-1)^{\nu_{\mathbf{q}}}$, as illustrated in Table~\ref{Tab:C2Z}.
Hence, the first transformation $\mathcal{U}_{\mathrm{I}}$ connects states $\ket{\mathbf{q}}$ and $\ket{R_{\mathbf{q}}}$. 
The second transformation $\mathcal{U}_{\mathrm{II}}$ reverses the dynamics due to $\mathcal{U}_{\mathrm{I}}$ and returns the system to its original state, up to the geometric phase 
\begin{equation}
    \phi_g = \nu_{\mathbf{q}} \pi \mod2\pi. \label{eq:phigeom}
\end{equation}

The combined transformation $\ket{\mathbf{q}} \rightarrow \mathcal{U}_{\mathrm{II}}\mathcal{U}_{\mathrm{I}}\ket{\mathbf{q}}=e^{i\nu_{\mathbf{q}} \pi}\ket{\mathbf{q}}$ is, up to single-qubit rotations, equivalent to the C$_k$Z gate. 
The full circuit implementing the C$_k$Z gate thus consists of applying the X and Z gates to all but one qubit before $\mathcal{U}_{\mathrm{I}}$ and $\mathcal{U}_{\mathrm{II}}$ followed by applying the X gates to the same qubits, as illustrated in Fig.~\ref{Fig:XZXGate_protocol} and detailed in Appendix \ref{App:Correction_Circuit}.

%%%%%%%%%%%%%%%%%%%%%%%%%
\begin{figure}[b]
\centering
\includegraphics[width=0.9\columnwidth]{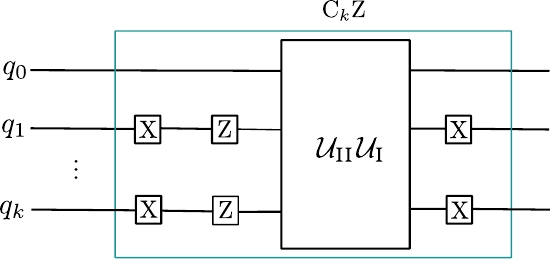} 
\caption{Circuit implementing the C$_k$Z gate. Applying the Hadamard gates on the target qubit before and after the C$_k$Z gate will result in C$_k$NOT, which is the Toffoli gate for $k=2$.
}
\label{Fig:XZXGate_protocol}
\end{figure}
%%%%%%%%%%%%%%%%%%%%%%%%%%%%

\subsection{Adiabatic dynamics of the system} \label{Subseq:Microscopic_dynamics}

\begin{figure*}[t]
\includegraphics[width=\textwidth]{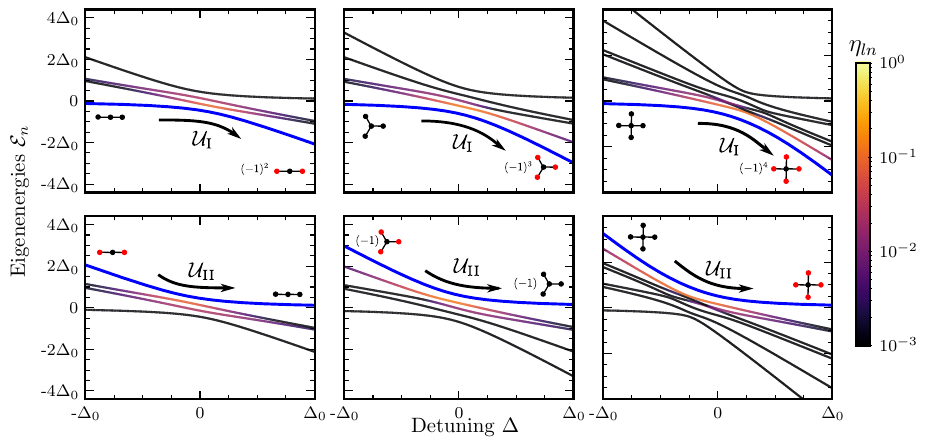}
\caption{Energy eigenvalues $\mathcal{E}_n$ of Hamiltonian (\ref{Eq:Hamiltonian}) for $N=3,4,5$ atoms (left, center, right panels) arranged in symmetric star-graph configurations 
($B_{0j}= B, \, B_{jj+1}= \mathrm{const} \, \forall \, j \geq 1$) vs detuning $\Delta$.
The parameters are $\Omega=\Omega_0 \, \forall \, \Delta \in [-\Delta_0, \Delta_0]$; 
$\Delta_0=2.4\Omega_0$, $|B|=6\Omega_0$ ($B\gtrless 0$ for upper/lower panels) for $N=3,4$; 
and $\Delta_0=3.2\Omega_0$, $|B|=5.6\Omega_0$ for $N=5$.
In step I (upper panels) we implement the transformation $\mathcal{U}_{\mathrm{I}}$: 
Starting in state $\ket{\mathbf{q}} = \ket{\alpha_1}$ for $\Delta=-\Delta_0$, the system adiabatically follows the eigenstate $\ket{\alpha_1}$ with lowest energy $\mathcal{E}_1$ (solid blue line) reaching state $\ket{R_\mathbf{q}}$ for $\Delta=\Delta_0$. 
 In step II (lower panels) we implement the transformation $\mathcal{U}_{\mathrm{II}}$:  
Now starting in state $\ket{R_\mathbf{q}} = \ket{\alpha_m}$ for $\Delta=-\Delta_0$, the system adiabatically follows the eigenstate $\ket{\alpha_m}$ with highest energy $\mathcal{E}_m$ (solid blue line) returning to state $\ket{\mathbf{q}}$ at $\Delta=\Delta_0$ and acquiring the sign change $(-1)^{\nu_{\mathbf{q}}}$.
During steps I and II, the non-adiabatic transitions away from states $\ket{\alpha_{l=1,m}}$ to other instantaneous eigenstates $\ket{\alpha_n}$ are quantified by the dimensionless parameter $\eta_{ln}=|\langle \alpha_l|\partial_t|\alpha_n\rangle|^2 \tau/\Delta_0$ (color depth of the solid lines for the corresponding $\mathcal{E}_n$).
Eigenenergies of ``dark'' eigenstates, decoupled from the laser for all $\Delta$, are not shown. }
\label{Fig:Spectrum}
\end{figure*}

The instantaneous eigenstates $\ket{\alpha_n}$ of Hamiltonian (\ref{Eq:Hamiltonian}) for any $[\Omega, \Delta, B]$ are defined via
\begin{equation}
\mathcal{H}(t)\ket{\alpha_n(t)}=\mathcal{E}_n(t)\ket{\alpha_n(t)} , \label{Eq:AdiabatSpectrum}
\end{equation}
where $\mathcal{E}_n$ are the corresponding energy eigenvalues. 
Consider the low-energy part of the spectrum $|\mathcal{E}_n| < |B|$ not suppressed by the blockade interactions $B$, see Fig.~\ref{Fig:Spectrum}. 
There are $m=2^k+1$ states in this subspace. 
Applying the parity operator $\mathcal{P}$ to both sides of Eq.~(\ref{Eq:AdiabatSpectrum}) and using the equality $\mathcal{P}\mathcal{H}[\Omega,\Delta,B]=-\mathcal{H}[\Omega,-\Delta,-B]\mathcal{P}$, we find that 
\begin{equation} \label{Eq:Symemtry_Spectrum}
    \mathcal{E}_n[\Omega,\Delta,B]=-\mathcal{E}_{n}'[\Omega,-\Delta,-B],
\end{equation}
i.e., for every eigenstate $\ket{\alpha_n}$ of $\mathcal{H} [\Omega, \Delta, B]$ with eigenvalue $\mathcal{E}_n [\Omega, \Delta, B]$, there is an eigenstate $\ket{\alpha_n'} = \mathcal{P}\ket{\alpha_n}$ of $\mathcal{H} [\Omega, -\Delta, -B]$ with eigenvalue $\mathcal{E}_n' [\Omega, -\Delta,-B] = - \mathcal{E}_n [\Omega, \Delta,B]$.

%%%%%%%%%%%%%%%%%%%%%%%%%%%%%%%%%%%%%%%%%%%%%%%%%%%
%%%%%%%%%%%%%%%%%%%%%%%%%%%%%%%%%%%%%%%%%%%%%%%%%

Recall now that, for large negative detuning $-\Delta \gg \Omega$, the lowest energy eigenstate $\ket{\alpha_1} = \ket{\mathbf{q}}$ with $\mathcal{E}_1 \simeq 0$ coincides with the state $\ket{\mathbf{q}}$ with all the atoms in the ground states, whereas for large positive detuning $\Delta \gg \Omega$ the same eigenstate $\ket{\alpha_1} = \ket{R_\mathbf{q}}$ with energy $\mathcal{E}_1 \simeq -\nu_{\mathbf{q}} \Delta_0$ coincides with the state $\ket{R_\mathbf{q}}$ with $\nu_{\mathbf{q}}$ Rydberg excitations. 
Thus, in step I, starting from state $\ket{\mathbf{q}}$ and assuming $B>0$, we turn on $\Omega(t)$ for time $0 \leq t \leq \tau$ and simultaneously sweep the laser detuning $\Delta(t)$ from a large negative value $-\Delta_0$ to a large positive value $\Delta_0 <B$, which implements the transformation 
\begin{equation}
    \mathcal{U}_{\mathrm{I}}=
    \exp \left[-i\int_0^\tau \! \mathcal{H}(t)dt \right] =
    \mathcal{U} [\Omega(t), \Delta(t), B] . \label{eq:UI}
\end{equation}
For sufficiently slow sweep (see below), 
%$\eta_{1n}=|\langle \alpha_1|\partial_t|\alpha_n\rangle^2 T/\Delta_0 \ll 1 \, \forall \, n=2,3, \ldots , m$, 
the system adiabatically follows the eigenstate $\ket{\alpha_1}$ and transitions from $\ket{\mathbf{q}}$ to  $e^{i \phi_d^{\mathrm{I}}}\ket{R_{\mathbf{q}}}$, where $\phi_d^{\mathrm{I}}=-\int_0^\tau \mathcal{E}_1(t) dt$ is the acquired dynamical phase.

Next, in step II, we set $B<0$ (see below), switch on the laser and sweep its detuning as
\begin{equation}
\Omega(t) = \Omega(2\tau-t), \quad \Delta(t) = - \Delta(2\tau-t) \label{eq:OmegaDelta}
\end{equation}
for time $\tau<t \leq 2\tau$. 
This implements the transformation
\begin{align}
  \mathcal{U}_{\mathrm{II}} &= \exp \left[-i\int_\tau^{2\tau} \!\! \mathcal{H}(t)dt \right] = \mathcal{U} [\Omega(t), \Delta(t), -B] \nonumber \\
  &= 
    \mathcal{P}\mathcal{U}_{\mathrm{I}}^{-1}\mathcal{P} . \label{eq:UII}
\end{align}
Now for $\Delta = -\Delta_0$ the highest energy eigenstate $\ket{\alpha_m} = \ket{R_{\mathbf{q}}}$ with $\mathcal{E}_m \simeq \nu_{\mathbf{q}} \Delta_0$ coincides with the state $\ket{R_{\mathbf{q}}}$, whereas for $\Delta =\Delta_0$ the same eigenstate $\ket{\alpha_m} = \ket{\mathbf{q}}$ with energy $\mathcal{E}_m \simeq 0$ coincides with state $\ket{\mathbf{q}}$ with all the atoms in the ground states. 
Hence, 
%for $\eta_{mn}=|\langle \alpha_m|\partial_t|\alpha_n\rangle^2 T/\Delta_0 \ll 1 \, \forall \, n=m-1,m-2, \ldots , 1$, 
the system adiabatically following $\ket{\alpha_m}$ transitions from $\ket{R_{\mathbf{q}}}$ to $e^{i \phi_d^{\mathrm{II}}}\ket{\mathbf{q}}$, where $\phi_d^{\mathrm{II}}=-\int_\tau^{2\tau} \mathcal{E}_m(t) dt$ is the dynamical phase. 
Due to the symmetry of the spectrum in Eq.~(\ref{Eq:Symemtry_Spectrum}) and the temporal profile of the laser pulses in Eq.~(\ref{eq:OmegaDelta}), the dynamical phases $\phi_d^{\mathrm{I}} = - \phi_d^{\mathrm{II}}$ exactly cancel, and only the geometric phase $\phi_g$ in Eq.~(\ref{eq:phigeom}) remains. 

%%%%%%%%%%%%%%%%%%%%%%%%%%%%%%%%%%%%%%%%%%%%%%%
\begin{figure}[t]
\includegraphics[width=1.0\columnwidth]{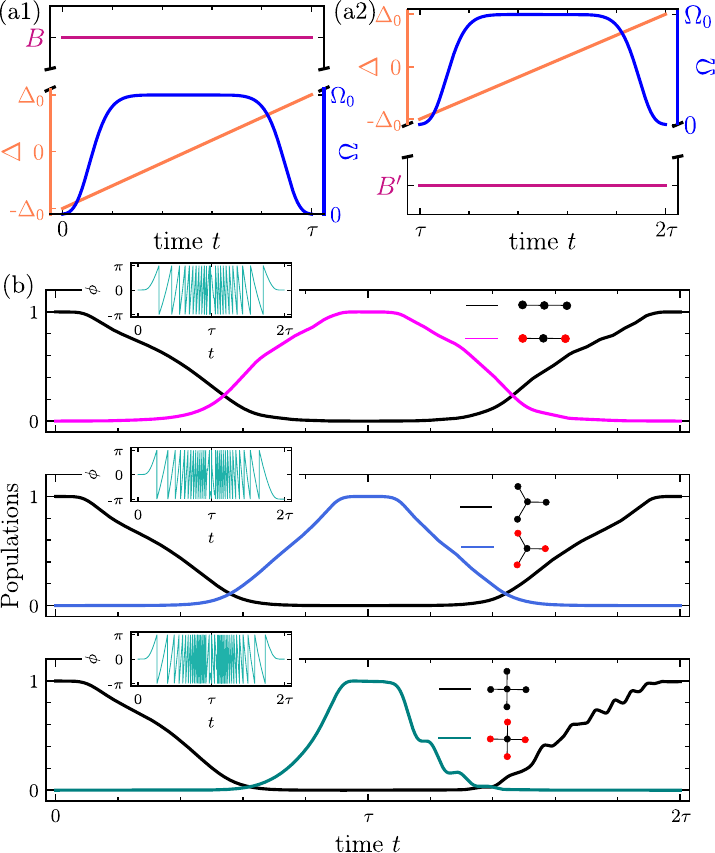} 
\caption{(a) Time dependence of the Rabi frequency $\Omega(t)$ (right vertical axis) and detuning $\Delta(t)$ (left vertical axis) during step I with $B>0$ (a1) and step II with $B<0$ (a2) as per Eqs.~(\ref{eq:PulseI}) and (\ref{eq:OmegaDelta}), with parameters $\Omega_0,\Delta_0,B$ as in Fig. \ref{Fig:Spectrum}.
(b) Dynamics of populations of the initial/final states $\ket{\mathbf{q}}=\ket{11\ldots 1}$ and the corresponding intermediate MIS states $\ket{R_{\mathbf{q}}}=\ket{1r\ldots r}$ with $\nu_{\mathbf{q}} = k$ Rydberg excitations for $N=k+1=3,4,5$ atom graphs (top, middle, bottom panels) 
with $\tau=16\pi/\Omega_0$. 
Inset in each panel shows the total phase $\phi(t) = \arg \braket{\Psi(0)}{\Psi(t)}$ of the state $\ket{\Psi}$ of the system.}
\label{Fig:TrDynamics}
\end{figure}
%%%%%%%%%%%%%%%%%%%%%%%%%%%%%%%%%%%%%%%%%%%%%%%

To illustrate the foregoing discussion, in Fig.~\ref{Fig:Spectrum} we show the adiabatic spectrum of the system with $N=3,4,5$ atoms in symmetric star-graph configurations and the adiabatic paths that the system takes during steps I and II. 
We simulate the dynamics of the system driven by flat-top laser pulses with linear sweep of the detuning satisfying Eq.~(\ref{eq:OmegaDelta}), as shown in Fig.~\ref{Fig:TrDynamics}(a). 
The corresponding dynamics are illustrated in  Fig.~\ref{Fig:TrDynamics}(b), while in Appendix~\ref{App:Population_phase_dynamics} we show the dynamics of the system for all the possible inputs implementing the C$_k$Z gates with $k=2,3,4$.
To insure the symmetry of the spectrum $\mathcal{E}_1[\Omega,\Delta,B] = -\mathcal{E}_m[\Omega,-\Delta,-B]$ leading to exact cancellation of the dynamical phases $\phi_d^{\mathrm{I}} + \phi_d^{\mathrm{II}} =0$, we assume that the sign of the interaction $B\to -B = B'$ can be changed between the two pulses by quickly transferring the atoms to another Rydberg state $\ket{r} \to \ket{r'}$, as discussed in Sec.~\ref{sec:BtomB} and Appendix~\ref{app:BtomB}.  

In order for the system to adiabatically follow the instantaneous eigenstates $\ket{\alpha_l(t)}$ during steps I ($l=1$) and II ($l=m$), the non-adiabatic transition rates $\bra{\alpha_l(t)} \partial_t \ket{\alpha_n(t)}=\bra{\alpha_l} \partial_t \mathcal{H}\ket{\alpha_n}/(\mathcal{E}_l -\mathcal{E}_n)$ to all the other eigenstates $\ket{\alpha_n(t)}$ should be small compared to their energy separation $|\mathcal{E}_l - \mathcal{E}_n|$.  
For the star-graph configurations, the lowest and highest energy eigenstates $\ket{\alpha_{1,m}}$ remain gapped for all $\Delta$, with the minimal gap $\delta \mathcal{E} = \min |\mathcal{E}_1 - \mathcal{E}_2| = \min |\mathcal{E}_m - \mathcal{E}_{m-1}| \propto \Omega_0$ around $\Delta=0$. 
In Fig.~\ref{Fig:Spectrum}, we quantify the transition rates from states $\ket{\alpha_{l=1,m}}$ by the dimensionless parameter $\eta_{ln}=|\langle \alpha_l|\partial_t|\alpha_n\rangle|^2\tau/\Delta_0$, smallness of which in the vicinity of $\Delta(t)=0$ ($t=\tau/2$ and $t=3\tau/2$) ensures good adiabatic following. 
We note finally that only symmetric states are coupled to the ground state by the laser acting uniformly on all the atoms, while non-symmetric states remain ``dark'', $\eta_{ln}=0$, for all times $t$ and any $2\Delta_0/\tau=\beta$, as discussed in Appendix \ref{Appendix:Dark_states}.

\section{Gate Fidelity} \label{Sec:Fidelity}

We quantify the performance of our protocol by error probability $E=1-F$, where $F$ is the gate fidelity  averaged over all the $d=2^N$ input states of $N=k+1$ qubits \cite{pedersen2007fidelity} 
\begin{equation}
    F=\frac{1}{d(d+1)}\big(\Tr{\mathcal{M}\mathcal{M}^{\dagger}}+|\Tr{\mathcal{M}}|^2\big),
\end{equation}
with $\mathcal{M}=\mathcal{U}^{\dagger}_{\text{C}_{k}\text{Z}} \tilde{\mathcal{U}}$, where $\mathcal{U}_{\text{C}_{k}\text{Z}}$ is the unitary transformation corresponding to the ideal C$_k$Z gate, and $\tilde{\mathcal{U}}$ is the actual transformation. 
To determine $\tilde{\mathcal{U}}$, we solve the Schrödinger equation $i\partial_t \ket{\Psi} = \tilde{\mathcal{H}} \ket{\Psi}$ for the system initially in each of the input states $\ket{\Psi(0)}=\ket{\mathbf{q}}=\ket{q_0q_1...q_k}$ using the effective non-Hermitian Hamiltonian 
\begin{align} \label{Eq:Non_Herm_Hamiltonian}
    \tilde{\mathcal{H}}(t)&=\mathcal{H}(t)-\frac{i}{2}\sum_{i=0}^{k} \big[ \Gamma_r \ketbra{r_i}{r_i} + \Gamma_{r'} \ketbra{r_i'}{r_i'} \big] , 
\end{align}
where $\Gamma_{r,r'}$ are the decay rates of the Rydberg states $\ket{r},\ket{r'}$ to states outside the computational subspace.  
%This decay reduces the norm of $\ket{\Psi(t)}$. 
After the evolution, we project $\ket{\Psi(2\tau)}$ onto the computational subspace, $\ket{\tilde{\Psi}} = \Pi_{\mathbf{q}}\ket{\Psi(2\tau)}$,  $\Pi_{\mathbf{q}} = \sum_{i=0}^k \sum_{q=0,1} \ket{q_i}\bra{q_i}$, and obtain from $\ket{\tilde{\Psi}}$ the transformation matrix $\tilde{\mathcal{U}}$, which is diagonal since $\mathcal{H}$ does not couple the qubits states $\ket{0,1}$, and non-unitary due to the decay of the Rydberg states and projection onto the computational basis. 
Neglecting the decay back to the qubit states slightly overestimates the gate error but simplifies the calculations. 

\subsection{Decay and leakage errors} 
\label{sec:ErrorDL}

The two main sources of error in our protocol are the decay of the Rydberg states and non-adiabatic transitions to the states with the wrong parity. 

The probability of decay during steps I and II is 
\begin{equation}
E_{\textrm{decay}} = 1 - \frac{1}{2^{k+1}}\sum_{\mathbf{q}}e^{ - \Gamma \int_0^{2\tau} \nu_{\mathbf{q}}(t) dt } \simeq 2 \bar{\nu}_k \Gamma \tau, \label{eq:Edecay}
\end{equation}
where $\bar{\nu}_k = \tfrac{1}{2^{k+1}} \sum_{\mathbf{q}} \bar{\nu}_{\mathbf{q}}$ is the mean number of Rydberg excitations averaged over all the input configurations $\ket{\mathbf{q}}$ and pulse duration $\tau$, $\bar{\nu}_{\mathbf{q}} = \tfrac{1}{\tau} \int_0^{\tau} \nu_{\mathbf{q}}(t) dt$, 
and we assume that the atoms in Rydberg states $\ket{r}, \ket{r'}$ decay with rate $\Gamma_{r,r'} \simeq \Gamma$ and neglect the transfer time between these states discussed in Sec.~\ref{sec:BtomB} and Appendix~\ref{app:BtomB}.
Note that at time $t=\tau$, the input-averaged number of Rydberg excitations is 
$\nu_k = \tfrac{1}{2^{k+1}} \sum_{\mathbf{q}} \nu_{\mathbf{q}}(\tau) = 9/8, 25/16, 65/32$ for $k=2,3,4$ ($N=3,4,5$), and if we assumed that the atoms spend on average half of the total time $2\tau$ in the Rydberg states $\ket{r},\ket{r'}$, we could approximate $\bar{\nu}_k \approx \tfrac{1}{2}\nu_k$, which, however, would slightly overestimate the decay errors for $k=3$ and even more so for $k=4$.    
Actually, due to the interaction between the outer atoms, $B_{ij} =B/64,B/27,B/8$ for $k=2,3,4$, their transition from state $\ket{1}$ to the Rydberg state $\ket{r}$ occurs around slightly later times $t> \tau/2$ when $\Delta(t) \sim B_{ij}>0$ during step I; similarly, during step II, they transition from state $\ket{r'}$ back to state $\ket{1}$ slightly earlier, around time $t<3\tau/2$ when $\Delta(t) \sim B'_{ij}<0$.
Hence the time-averaged $\bar{\nu}_{\mathbf{q}}$ are slightly smaller than $\frac{1}{2} \nu_{\mathbf{q}}$, but we can estimate $\bar{\nu}_{\mathbf{q}}$ using the mean-field approximation. 
We thus assume that each outer atom $i$ initially in state $\ket{1_i}$ is in the field $B_i^{(\mathbf{q})} \simeq \sum_{j\neq i} B_{ij} \braket{R_{\mathbf{q}}}{r_j} \braket{r_j}{R_{\mathbf{q}}}$  produced by the other atoms $j$ in Rydberg state $\ket{r_j}$ as per state $\ket{R_{\mathbf{q}}}$ and therefore transitions to the Rydberg state around time $t>\tau/2$ when $\Delta(t) = \frac{1}{2} B_i^{(\mathbf{q})}$. 
We thus obtain $2\bar{\nu}_k = 8.9/8, 23.9/16, 57.4/32$ for $k=2,3,4$. 

\begin{figure}[t!]
\centering  
\includegraphics[width=1.0\columnwidth]{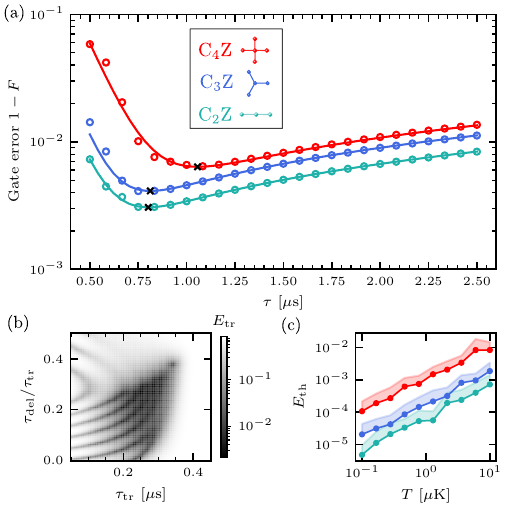} 
\caption{(a) Total error probability (infidelity) $E=1-F = E_{\text{decay}} + E_{\text{leakage}}$ of gate C$_k$Z ($k=2,3,4$) vs pulse duration $\tau$, as obtained  numerically (open circles) using the full Hamiltonian (\ref{Eq:Non_Herm_Hamiltonian}) and analytically using Eq.~(\ref{eq:Edecay}) with the mean excitation numbers $2\bar{\nu}_{2,3,4} \simeq (1.11, 1.49, 1.79)$  
and Eq.~(\ref{eq:Eleak_fit}) with the fitting parameters $\mu_{2,3,4} = (0.99,4.27,15.88)$ and $c_{2,3,4}=(0.43, 0.45, 0.48 )$. 
The optimal pulse duration $\tau_{\text{opt}}$ and corresponding minimum error $E_{\text{min}}$ (black crosses) for each gate are obtained using Eqs. \eqref{Eq:Optimal_time} and \eqref{eq:Emin} respectively. Other parameters are $\Omega_0=2\pi \times 8\:$MHz, $\Gamma= 2\pi \times 0.5\,$kHz for all $k$, $\Delta_0=2.4\Omega_0$ and $|B|=6\Omega_0$ for $k=2,3$, and $\Delta_0= 3.2\Omega_0$ and $|B|=5.6\Omega_0$ for $k=4$. 
(b) Error $E_{\mathrm{tr}}$ for the STIRAP transfer $\ket{r} \to \ket{r'}$ via  intermediate state $\ket{p}$ using two laser pulses $\Omega_{SP,DP}(t) = \mp \Omega_{\max} \sin^2[\pi (t-t_{SP,DP})/(\tau_{\mathrm{tr}} - \tau_{\mathrm{del}})]$ with peak Rabi frequencies $\Omega_{\max} = 2\pi \times 80\:$MHz vs the total transfer time $\tau_{\mathrm{tr}}$ and pulse delay time $\tau_{\mathrm{del}}=t_{SP}-t_{DP}$.
(c) Thermal dephasing error $E_{\mathrm{th}}$ for the C$_k$Z ($k=2,3,4$) gates vs temperature $T$ as obtained via simulations of the total process for all the input states. 
Data point for each $T$ (filled circles) is obtained by averaging over $M=40$ independent atomic velocity sets sampled from the Maxwell-Boltzmann distribution, with the shading corresponding to one standard deviation. } 
\label{Fig:Fidelity_345}
\end{figure}

At the end of the process, the probability of error due to nonadiabatic transfer, or leakage, to the states with the wrong parity is given by 
\begin{equation}
    E_{\text{leakage}}=\frac{4}{2^{k+1}}\sum_{\mathbf{q}}|b_{\mathbf{q}}|^2,  \label{eq:Eleak}
\end{equation}
where, for each configuration $\ket{\mathbf{q}}$, $|b_{\mathbf{q}}|^2$ is the population transferred from state $\ket{\alpha_1}$ to the states with $\nu_{\mathbf{q}} \pm 1$ (or $\nu_\mathbf{q}\pm 3$ etc.) Rydberg excitations, and we assume that the population  $|b_{\mathbf{q}}|^2$ leaked from state $\ket{\alpha_1}$ at the end of step I is the same as that leaked from state $\ket{\alpha_m}$ at the end of step II, see Appendix~\ref{App:Non-adiabatic-trans}.   
Most of the leakage occurs to the nearest ``bright'' state $\ket{\alpha_{2}}$ with one less Rydberg excitation, $\nu_\mathbf{q} - 1$, and we estimate the non-adiabatic transition probability to that state using the Landau-Zener formula 
\cite{tzortzakakis2022microscopic}
\[
|b_{\mathbf{q}}|^2 \approx \exp \left[ -2\pi\frac{(\delta\mathcal{E}/2)^2}{2\Delta_0/\tau}\right] ,
\]
where $\delta \mathcal{E} = \min |\mathcal{E}_1- \mathcal{E}_2|$ is the minimal energy gap between states $\ket{\alpha_{1}}$ and $\ket{\alpha_{2}}$ in the vicinity of $\Delta=0$. 
The gap depends on the initial configuration $\ket{\mathbf{q}}$, and for each $k=2,3,4$ ($N=k+1$) the configurations with the smallest energy gap $\delta \mathcal{E}_k =g_k\Omega_0$ give the dominant contribution to the error in Eq.~(\ref{eq:Eleak}). 
Thus, for $k=2$, we have one ($\mu_{k=2}=1$) configuration, $\ket{111}$, with the smallest gap $\delta \mathcal{E}_2$   \cite{tzortzakakis2022microscopic}, while for $k=3,4$, we have $\mu_{3,4}=4,16$ configurations with the smallest gap $\delta \mathcal{E}_{3,4}$, 
as detailed in Appendix~\ref{App:Non-adiabatic-trans}.  
We can thus approximate Eq.~(\ref{eq:Eleak}) as 
\begin{equation} \label{eq:Eleak_fit}
    E_{\text{leakage}} \approx \frac{4}{2^{k+1}}\mu_k \exp \left[ -c_k\frac{\Omega_0^2}{\Delta_0}\tau \right],
\end{equation}
where $c_k=\pi g_k^2/4$. 
We verified these approximations by solving the Schrödinger equation with the full Hermitian Hamiltonian $\mathcal{H}(t)$ (without the atomic decay) for all $k$ and $\ket{\mathbf{q}}$ and fitting Eq.~(\ref{eq:Eleak_fit}), obtaining the values of $\mu_k$ and $c_k$ (or $g_k=2\sqrt{c_k/\pi}$) listed in the caption of Fig.~\ref{Fig:Fidelity_345}. 

%%%%%%%%%%%%%%%%%%%%%%%%%%%%%%%%%%%%%%%%%%%
%%%%%%%%%%%%%%%%%%%%%%%%%%%%%%%%%%%%

Adding now the decay of the Rydberg states, we obtain the total gate error $E=E_{\text{decay}} + E_{\text{leakage}}$ shown in Fig.~\ref{Fig:Fidelity_345}(a) for different pulse durations $\tau$.
Clearly, shorter pulses lead to smaller decay error, while longer pulses reduce the non-adiabatic leakage errors. 
To determine the optimal pulse duration $\tau_{\text{opt}}$, for each $k$, we minimize the total error $E$ and obtain 
\begin{equation} \label{Eq:Optimal_time}
\tau_{\text{opt}}\simeq \frac{\Delta_0}{c_k\Omega_0^2}\ln(\frac{\mu_k c_k\Omega_0^2}{2^{k}\bar{\nu}_k\Gamma\Delta_0}) ,
\end{equation}
leading to minimal error
\begin{equation} 
E_{\text{min}} \simeq 
2\bar{\nu}_k\Gamma \left[\tau_{\text{opt}} + \frac{\Delta_0}{c_k\Omega_0^2} \right] , \label{eq:Emin}
\end{equation}
which is in very good agreement with the exact numerical calculations for the full system.

\subsection{Rydberg state transfer error} 
\label{sec:BtomB}

As mentioned above, switching the sign of interaction $B\to -B = B'$ between the steps I and II can be accomplished by quickly transferring the atoms from Rydberg state $\ket{r}=\ket{n_S S_{1/2}}$ with $C_6>0$ to another Rydberg state $\ket{r'} = \ket{n_D D_{5/2}}$ with $C_6' \simeq -C_6$, and we assume Rb atoms.
This can be accomplished with high fidelity by stimulated Raman adiabatic transfer (STIRAP) using a pair of laser pulses $\Omega_{SP}(t)$ and $\Omega_{DP}(t)$ coupling near-resonantly the Rydberg states $\ket{n_S S_{1/2}}$ and $\ket{n_D D_{5/2}}$ to the intermediate state $\ket{6 P_{3/2}}$, see Appendix~\ref{app:BtomB}.
Applying the pulses in the ``counter-intuitive'' order, with $\Omega_{SP}(t)$ delayed from $\Omega_{DP}(t)$ by time $\tau_{\mathrm{del}}$ while still having sufficient temporal overlap, we scan $\tau_{\mathrm{del}}<\tau_{\mathrm{tr}}$ and the total transfer duration $\tau_{\mathrm{tr}}$ to find their optimal values and the corresponding transfer error.
For short times, this error stems mainly from the decay of the intermediate state $\ket{6 P_{3/2}}$ which is, however, highly suppressed when STIRAP is adiabatic, requiring laser pulses with large area $\int \bar{\Omega}(t) dt \gtrsim 2 \pi \times 10$, $\bar{\Omega}(t) \equiv \sqrt{|\Omega_{SP}(t)|^2 + |\Omega_{DP}(t)|^2}$  \cite{RevModPhys.70.1003,RevModPhys.89.015006}. 
For longer durations with good adiabatic following of the ``dark'' state $\ket{\psi_0} \propto \Omega_{DP} \ket{r} - \Omega_{SP} \ket{r'}$, the error is dominated by the decay of the Rydberg states. 
Remarkably, each Rydberg atom adiabatically follows the state $\ket{\psi_0}$ even in the presence of Rydberg-state interactions with other atoms leading to Raman detuning $\Delta_R \lesssim B_{ij} - B'_{ij}$, $i,j\geq 1$, provided $|\Delta_R(t)| \ll \bar{\Omega}(t)$.  

In Fig.~\ref{Fig:Fidelity_345}(b) we show the resulting transfer error $E_{\mathrm{tr}}$ averaged over all the input states for the C$_2$Z gate. 
For realistic parameters and total duration of the two pulses $\tau_{\mathrm{tr}}=0.2-0.3\:\mu$s and optimal delay $\tau_{\mathrm{del}} \sim 0.2 \tau_{\mathrm{tr}}$, we obtain the error $E_{\mathrm{tr}} \lesssim 2\times 10^{-3}$, which is smaller than the minimal total error $E_\mathrm{min}$ in Eq.~(\ref{eq:Emin}) and Fig.~\ref{Fig:Fidelity_345}(a). 
We obtain that $E_{\mathrm{tr}} < E_\mathrm{min}$ also for the C$_3$Z and C$_4$Z gates:
$E_{\mathrm{tr}}=(3.3,9.3) \times 10^{-3}$ with $\tau_{\mathrm{tr}}=(0.3,0.23) \mu$s for $k=3,4$, respectively. 

\subsection{Error due to thermal motion}  \label{Subsec:Thermal}

By changing the sign of interatomic interactions between steps I and II, $B_{ij}' = -B_{ij} \: \forall \: i,j$, we satisfy the condition for the symmetry of the spectrum in Eq.~(\ref{Eq:Symemtry_Spectrum}) which ensures that the dynamics of the system during step II is the time-reversal of the dynamics in step I, up to the geometric phase inherent from the parity transformation, cf. Eqs.~(\ref{eq:UI}) and (\ref{eq:UII}). 
Hence, static disorder of atomic positions, and the resulting variations in the interatomic interactions $B_{ij}^{(\prime)}$, do not affect the gate fidelity, provided the conditions $|B_{0j}^{(\prime)}|\gg\max(\Delta,\Omega)$, $|B_{ij}^{(\prime)}| \ll\max(\Omega) \: \forall \: i,j>0$ remain satisfied. 
But if the atoms move during and between steps I and II, the interactions will not satisfy the condition $B_{ij}' = -B_{ij}$ and the dynamical phases will not fully cancel, leading to dephasing. 

We thus need to consider the influence of thermal atomic motion on the gate fidelity. 
For pairs of atoms initially at positions $\mathbf{x}_{i,j}^{(0)}$ and moving with velocities $\mathbf{v}_{i,j}$, the interaction strength varies with time as $B_{ij}^{(\prime)}(t)=C_6^{(\prime)} \left| \mathbf{x}_{i}^{(0)}-\mathbf{x}_{j}^{(0)} + (\mathbf{v}_i - \mathbf{v}_j)t \right|^{-6}$. 
At temperature $T$, the atomic velocities are given by the Maxwell-Boltzmann distribution $W(\mathbf{v}) = \left( \sqrt{\pi} u \right)^{-3} \exp(-\mathbf{v}^2/u^2)$, where $u=\sqrt{2k_{\mathrm{B}} T/m_{\mathrm{Rb}}}$ is the most probable velocity of the atoms, assumed $^{87}\text{Rb}$. 
Treating the atomic motion classically, we place each atom $i$ at position $\mathbf{x}_{i}^{(0)}$ near the corresponding trap minimum, and choose a random velocity $\mathbf{v}_i$ according to the probability distribution $W(\mathbf{v})$. 
We then simulate the complete gate protocol with the time-dependent interactions $B_{ij}^{(\prime)}(t)$, assuming that the velocities $\mathbf{v}_i$ remain constant during the gate time. 
This is justified by the fact that, for typical temperatures $T=0.1-10\:\mu$K, leading to $u =10^{-3} - 10^{-2}\:$m/s, the atoms move during the gate time $2\tau = 2\:\mu$s little compared to the trap size $a_0 =\sqrt{\hbar /(2m_{\mathrm{Rb}} \nu)} \lesssim 0.1 \: \mu$m, where $\nu =2\pi \times 10\;$kHz is a typical trap frequency.   

In Fig.~\ref{Fig:Fidelity_345}(c) we show the C$_k$Z gate errors $E_{\mathrm{th}}$ resulting from the thermal atomic motion as obtained from the numerical simulations for various temperatures $T$.
For each temperature, we sample $M=40$ atomic velocities, and compute the total error $E_{\mathrm{tot}}$ averaged over all the gate inputs $\ket{\mathbf{q}}$. 
The error associated with the thermal atomic motion is then obtained by subtracting the error $E=E_{T=0}$ of Sec.~\ref{sec:ErrorDL} for frozen atoms, $E_{\mathrm{th}} = E_{\mathrm{tot}} - E_{T=0}$. 
We observe that for the experimentally revenant temperatures $T \sim 1\:\mu$K, the error due the thermal dephasing is at least an order of magnitude smaller than the combined decay and leakage error $E$ of Sec.~\ref{sec:ErrorDL}. 

\subsection{Faster gate with optimized adiabatic pulse} \label{Subsec:Optimization}

\begin{figure}[t]  
\includegraphics[width=\columnwidth]{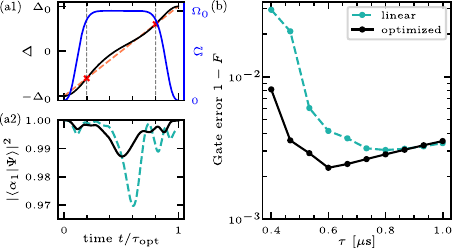} 
\caption{(a1) Time-dependence of the pulse Rabi frequency $\Omega(t)$ (solid blue line) and detuning $\Delta(t)$ for linear sweep (dashed orange line) and optimized adiabatic sweep (solid black line). Dashed vertical lines bound the interval $\delta t \leq t \leq \tau-\delta t$ where $\Omega(t) \geq 0.9 \Omega_0$ can be assumed approximately constant, with $\Delta(t)$ determined by Eq.~(\ref{eq:DeltaAdOpt}).
(a2) Dynamics of population of the adiabatic ground state $\ket{\alpha_1}$ for input $\ket{\mathbf{q}}=\ket{111}$, for the linear (dashed teal line) and optimized (solid black line) sweeps of $\Delta(t)$. 
(b) Total error probability $E=1-F = E_{\text{decay}} + E_{\text{leakage}}$ of the C$_2$Z gate vs pulse duration $\tau$ for the linear (dashed teal line) and optimized (black solid line) sweeps of $\Delta(t)$. 
For optimized $\Delta(t)$, the minimal gate error $\tilde{E}_{\text{min}}=2.3\cdot 10^{-3}=0.75 E_{\text{min}}$ is smaller, and is attained for shorter optimal pulse $\tilde{\tau}_{\text{opt}}=0.6\mu$s$=0.75\tau_{\text{opt}}$, than for the linear sweep of $\Delta(t)$, with all the parameters as in Fig.~\ref{Fig:Fidelity_345}(a) for $k=2$.}
\label{Fig:Opt_pulse}
\end{figure}

It is advantageous to perform faster gates with shorter pulses, but non-adiabatic transitions to the wrong parity states limit the maximal rate $\beta = 2\Delta_0/\tau$ of change of the detuning $\Delta(t)$. 
We can still reduce the pulse duration and increase the gate fidelity if, instead of the linear sweep of $\Delta(t)=\beta(t-\tau/2)$, we let $\dot{\Delta}(t)$ be determined by the energy gap $|\mathcal{E}_1 -\mathcal{E}_2|$ between the adiabatic ground $\ket{\alpha_1}$ and first excited $\ket{\alpha_2}$ states, during step I. 
[During step II, a mirror symmetric pulse satisfying Eq.~(\ref{eq:OmegaDelta}) will minimize nonadiabatic transitions between $\ket{\alpha_m}$ and $\ket{\alpha_{m-1}}$.]
Recall from Sec.~\ref{Subseq:Microscopic_dynamics} that, for constant $\Omega=\Omega_0$, the transition rate between the instantaneous eigenstates $\ket{\alpha_1}$ and $\ket{\alpha_2}$ can be cast as
\[
\eta_{12} = \frac{|\dot{\Delta}(t)|^2}{|\mathcal{E}_1(t) -\mathcal{E}_2(t)|^2} \frac{|\sum_i \braket{\alpha_1}{r_i} \braket{r_i}{\alpha_2}|^2}{\Delta_0/\tau} .
\]
Assuming that the numerator in the second fraction changes slowly
in the time interval $\delta t \leq t \leq \tau - \delta t$ where $\Omega(t) \approx \Omega_0$ can be assumed approximately constant, we determine the rate of change of the detuning via 
\begin{equation}
    \dot{\Delta}(t) = \zeta |\mathcal{E}_1(t) -\mathcal{E}_2(t)| , \quad  \zeta = \frac{2\Delta_0^*}{\int_{\delta t}^{\tau-\delta t} |\mathcal{E}_1 -\mathcal{E}_2|dt}
    \label{eq:DeltaAdOpt}
\end{equation}
in the corresponding detuning interval $-\Delta_0^* = \beta(\delta t -\tau/2) \leq \Delta \leq \Delta_0^* = \beta(\tau/2-\delta t)$; 
while in the boundary intervals $0 \leq t <\delta t$ and $\tau-\delta t \leq t \leq \tau$ where $\Omega(t)$ is changing rapidly, we interpolate between the points $\{(0,-\Delta_0), (\delta t, -\Delta_0^*)\}$ and $\{(\Delta_0^*,\tau - \delta t), (\tau, \Delta_0)\}$ using cubic polynomials with smooth derivatives $\dot{\Delta}(t)$ at $t=\delta t,\tau-\delta t$ and $\dot\Delta(0)=\dot\Delta(\tau)=0$, as illustrated Fig.~\ref{Fig:Opt_pulse}(a1). 
The resulting population dynamics of the instantaneous ground state $\ket{\alpha_1}$ for the three-atom initial state $\ket{\mathbf{q}} = \ket{111}$ is shown in Fig.~\ref{Fig:Opt_pulse}(a2) and compared to that for the linear sweep of $\Delta(t)$. 
We observe significant suppression of the non-adiabatic transition probability away from $\ket{\alpha_1}$ both during and, more importantly, at the end of the pulse.

%%%%%%%%%%%%%%%%%%%%%%%%%%%%%%%%%%%%%%%%%%%%%%%%%%
%%%%%%%%%%%%%%%%%%%%%%%%%%%%%%%%%%%%%%%%%%%

We use such pulses with optimized detuning $\Delta(t)$ to perform the C$_2$Z gate and calculate its fidelity averaged over all the input state while taking into account the Rydberg state decay, as in Fig.~\ref{Fig:Fidelity_345}(a). 
The results are shown in Fig.~\ref{Fig:Opt_pulse}, where we observe significant reduction of the total gate error $\tilde{E}_{\text{min}}=2.3\cdot 10^{-3}=0.75 E_{\text{min}}$ achieved for even shorter pulses of duration $\tilde{\tau}_{\text{opt}}=0.6\mu$s$=0.75\tau_{\text{opt}}$, as compared to the C$_2$Z gate performed with linearly chirped $\Delta(t)$. 

\section{Multi-qubit gates between distant atoms} \label{Sec:Distant_atoms}

%%%%%%%%%%%%%%%%%%%%%%%%%%%%%%%%%%%%%%%%%%%%%%%%%%
\begin{figure}[t]
\centering
\includegraphics[width=0.7\columnwidth]{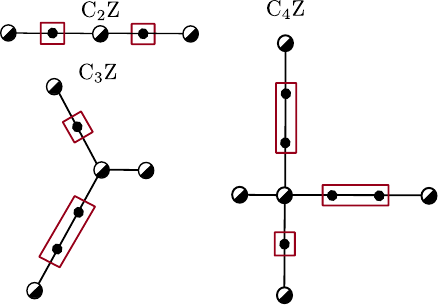} 
\caption{Examples of C$_2$Z, C$_3$Z, and C$_4$Z gates between distant atomic qubits. 
Auxiliary atoms inserted in each branch of the star-graph and initialized in state $\ket{1}$ (filled circles) play the role of a quantum bus mediating interactions between the distant qubit-encoding atoms (half-filled circles).}
\label{Fig:Distant_gates}
\end{figure}
%%%%%%%%%%%%%%%%%%%%%%%%%%%%%%%%%%%%%%%%%%%

We can adopt the approach of Ref. \cite{Doultsinos2025PRR} to implement quantum gates between distant atoms that do not directly interact with each other but are connected by auxiliary atoms playing the role of a quantum bus.
To this end, some or all of the outer atoms encoding qubits can be connected to the central atom $i=0$ by auxiliary atoms initialized in state $\ket{1}$, see Fig.~\ref{Fig:Distant_gates}. 
In each branch $i=1,2, \ldots , k$, we may have $n_i \geq 0$ auxiliary atoms mediating the interaction between the central atom and the outer atom.
The total number of atoms simultaneously addressed by the global laser pulses is now $N=k+1+\sum_{i=1}^k n_i$.
Denoting again the initial state of the central and outer atoms encoding qubits as $\ket{\mathbf{q}}=\ket{q_0q_1...q_k}$, $q_i=\{0,1\}$, the combined state of $N$ atoms can be cast as $\ket{\mathbf{Q}} \equiv \ket{\mathbf{q}} \prod_{i=1}^k \ket{1}^{\otimes n_i}$, where $\ket{1}^{\otimes n_i} = \ket{11\ldots 1}$ is the state of the quantum bus in $i$-th branch.
%which has $l_i = n_i +q_i$ atoms in state $\ket{1}$. 

%%%%%%%%%%%%%%%%%%%%%%%%%%%%%%%%%%%%%%%%%%%%%%%
\begin{table*}[t] 
    \caption{Illustration of $\mathcal{U}_{\mathrm{II}}\mathcal{U}_{\mathrm{I}}\ket{\mathbf{Q}}$ transformation for $N=5$ atoms, with the central and two end atoms encoding qubits while the auxiliary atoms are initialized in state $\ket{1}$. 
    The MIS state $\ket{R_\mathbf{Q}}$ for all inputs but $\ket{\mathbf{q}}=\ket{111}$ contains an even number of Rydberg excitations, $\tilde{\nu}_\mathbf{Q}=2$, which implements the C$_2$Z gate without additional single qubit operations. }
\centering
    \begin{tabular}{c c c c c }
        \hhline{=====}
        $\ket{q_0 q_1 q_2}$ & $\ket{\mathbf{Q}}$ & $\rightleftharpoons$ & $\ket{R_{\mathbf{Q}}}$ & Parity \\ 
        \hline
        $\ket{000}$ & $\circ$-$\bullet$-$\circ$-$\bullet$-$\circ$ & & $\circ$-$\textcolor{red}{\bullet}$-$\circ$-$\textcolor{red}{\bullet}$-$\circ$ & $+1$ \\
        $\ket{100}$ & $\circ$-$\bullet$-$\bullet$-$\bullet$-$\circ$ &  & $\circ$-$\textcolor{red}{\bullet}$-$\bullet$-$\textcolor{red}{\bullet}$-$\circ$ & $+1$ \\
        $\ket{010}$ & $\circ$-$\bullet$-$\circ$-$\bullet$-$\bullet$ &  & $\circ$-$\textcolor{red}{\bullet}$-$\circ$-$\bullet$-$\textcolor{red}{\bullet}$ & $+1$ \\
        $\ket{001}$ & $\bullet$-$\bullet$-$\circ$-$\bullet$-$\circ$ &  & $\textcolor{red}{\bullet}$-$\bullet$-$\circ$-$\textcolor{red}{\bullet}$-$\circ$ & $+1$ \\
        $\ket{110}$ & $\circ$-$\bullet$-$\bullet$-$\bullet$-$\bullet$ &  & $\frac{1}{\sqrt{2}}(\circ$-$\bullet$-$\textcolor{red}{\bullet}$-$\bullet$-$\textcolor{red}{\bullet}$ + $\circ$-$\textcolor{red}{\bullet}$-$\bullet$-$\textcolor{red}{\bullet}$-$\bullet$) & $+1$ \\
        $\ket{101}$ & $\bullet$-$\bullet$-$\bullet$-$\bullet$-$\circ$ &  & 
        $\frac{1}{\sqrt{2}}(\bullet$-$\textcolor{red}{\bullet}$-$\bullet$-$\textcolor{red}{\bullet}$-$\circ$ + $\textcolor{red}{\bullet}$-$\bullet$-$\textcolor{red}{\bullet}$-$\bullet$-$\circ$) & $+1$\\
        $\ket{011}$ & $\bullet$-$\bullet$-$\circ$-$\bullet$-$\bullet$ & & $\textcolor{red}{\bullet}$-$\bullet$-$\circ$-$\bullet$-$\textcolor{red}{\bullet}$ & $+1$ \\ 
        $\ket{111}$ & $\bullet$-$\bullet$-$\bullet$-$\bullet$-$\bullet$ & & $\textcolor{red}{\bullet}$-$\bullet$-$\textcolor{red}{\bullet}$-$\bullet$-$\textcolor{red}{\bullet}$  & $-1$ \\
        \hhline{=====}
    \end{tabular}
    \label{Tab:C2Zextend}
\end{table*}
%%%%%%%%%%%%%%%%%%%%%%%%%%%%%%%%%%%%%%%%%%%%%%%%%

For large negative detuning $-\Delta \gg \max (\Omega)$, state $\ket{\mathbf{Q}}$ is again the ground state of Hamiltonian (\ref{Eq:Hamiltonian}). 
Adiabatic sweep of the detuning $\Delta$ to a large positive value $\max (\Omega) \ll \Delta < B$, with $B$ being the vdW interaction between the nearest-neighbor atoms in Rydberg state $\ket{r}$,  we implement the transformation $\mathcal{U}_{\mathrm{I}}$ that prepares the system in the corresponding MIS state $\ket{R_{\mathbf{Q}}}$ with   
\begin{equation} \label{Eq:nu_r-extended}
\tilde{\nu}_\mathbf{Q} = 
\begin{cases}
1 + \sum_{i=1}^k\lceil l_i/2\rceil & \text{if }  q_0=1 \; \& \; l_i \text{ is even } \forall \, i \\
\sum_{i=1}^k\lceil l_i/2\rceil & \text{otherwise} 
\end{cases}
\end{equation}
Rydberg excitations, where $l_i = n_i+q_i$. 
We next switch the sign of interaction $B\to -B=B'$ by transferring the Rydberg atoms to state $\ket{R'_{\mathbf{Q}}}$ with all $\ket{r} \to \ket{r'}$. 
We then repeat the laser pulse to implement the reverse transformation $\mathcal{U}_{\mathrm{II}}$ which returns the system to the initial state $\ket{\mathbf{Q}}$ and imprints the geometric phase $\phi_g = \tilde{\nu}_\mathbf{Q} \pi \mod 2\pi$ stemming from the parity of $\tilde{\nu}_\mathbf{Q}$. 
This is illustrated in Table~\ref{Tab:C2Zextend} for $N=5$ atoms implementing the extended C$_2$Z gate. 
Depending on the number of auxiliary atoms in each branch of the graph, the combined transformation 
\begin{equation}
\mathcal{U}_{\mathrm{II}} \mathcal{U}_{\mathrm{I}} \ket{\mathbf{Q}} = (-1)^{\tilde{\nu}_\mathbf{Q}} \ket{\mathbf{Q}}
\end{equation}
is either equivalent to the C$_k$Z gate when $n_i$ is odd for all $i=1,2,\ldots, k$ (as in Table~\ref{Tab:C2Zextend}), or can be converted to the C$_k$Z gate using single-qubit operations X and Z, as discussed in Appendix~\ref{App:Correction_Circuit_extensions}. 

%%%%%%%%%%%%%%%%%%%%%%%%%%%%%%%%%%%%%%%%%%%
\begin{figure}[t] 
\includegraphics[width=\columnwidth]{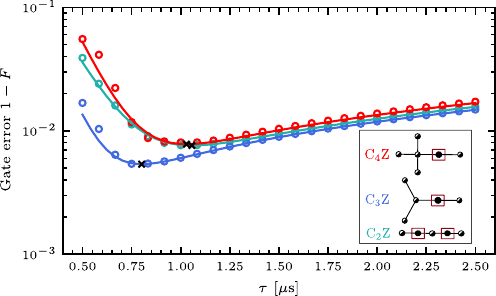} 
\caption{Total error probability (infidelity) $E=1-F = E_{\text{decay}} + E_{\text{leakage}}$ of gates C$_k$Z  ($k=2,3,4$) with extensions (auxiliary atoms) vs pulse duration $\tau$, as obtained  numerically (open circles) using the full Hamiltonian (\ref{Eq:Non_Herm_Hamiltonian}) and semi-analytically using Eq.~(\ref{eq:Edecay}) with $2\bar{\nu}_{2,3,4} \simeq (2.09,1.97,2.22)$ 
and Eq.~(\ref{eq:Eleak_fit}) with $\mu_{2,3,4} = (1.5,4.54, 17.45)$ and $c_{2,3,4}=(0.3, 0.45, 0.48)$. 
Parameters are as in Fig.~\ref{Fig:Fidelity_345}(a) for the corresponding $k$'s. 
For short pulse durations $\tau \lesssim 1\:\mu$s, the average error of the C$_2$Z gate with two auxiliary atoms is larger than that of the C$_3$Z and C$_4Z$ gates with one auxiliary atom, since for the former, when averaged over all inputs, more atoms initially in state $\ket{1}$ are participating in the dynamics, leading to significantly more leakage error.} 
\label{Fig:FidelityBus}
\end{figure}
%%%%%%%%%%%%%%%%%%%%%%%%%%%%%%%%%%%%

We performed numerical simulations of the complete protocol for the C$_k$Z gates including several auxiliary atoms.  
In Fig.~\ref{Fig:FidelityBus} we show the calculated average error probabilities for the the extended C$_k$Z gates involving one or two auxiliary atoms. 
We observe that gate fidelities are not adversely affected by the extensions and can still be sufficiently high even with $\sum_i n_i \lesssim 10$ auxiliary atoms. 

\section{Conclusions}
\label{Sec:Conclude}

To conclude, we have proposed and analyzed a protocol to perform efficient multiqubit C$_k$Z gates in quantum computing platform involving neutral atoms excited by the lasers to the strongly interacting Rydberg states. 
In our protocol, the atoms participating in the gate are arranged in star-graph configurations and addressed by global laser pulses that transfer them to the Rydberg state and back, leading to a conditional geometric phase $\pi$, while the dynamical phases cancel even in the presence of atomic position disorder at finite but sufficiently low temperature $T\lesssim 1\,\mu$K.  
We have also shown how to extend the effective interaction range to realize multiqubit gates between distant atoms, which will greatly improve connectivity in neutral atom quantum computers that rely on Rydberg blockade interactions.

We have gained detailed understating of adiabatic spectrum of the multiatom Rydberg system and non-adiabatic transitions that result in gate errors and should therefore be suppressed to achieve high-fidelity multiqubit quantum gates. 
We explored a simple method to reduce the harmful nonadiabatic transitions by adapting the sweeping rate of laser detuning to the energy gap between the instantaneous eigenstates of the Hamiltonian, which led to faster gate with better overall fidelity. 
More involved pulse optimization techniques \cite{cui2017optimal, Omran2019, jandura2022time, Doultsinos2025PRR} can yield even faster gates, but at the expense of greater sensitivity to uncertainty in interatomic distances and their interactions. 

Multiqubit gates between distant atoms can replace long sequences of universal two-qubits gates to perform desired unitary transformations during the execution of quantum algorithms \cite{martinez2016compiling}, and to realize error correction codes that encode individual logical qubits into several physical qubits and perform collective syndrome detection \cite{QCQI2000}.
Multiqubit gates can therefore reduce the total error while increasing the circuit depth and connectivity in neutral atom quantum computers that employ the Rydberg-state interactions. 
Thus, with realistic values of $\Gamma/B$ that we used, under otherwise optimal conditions \cite{jandura2022time} the best achieved or projected two-qubit  CZ (or CNOT) gate infidelities (errors) are $E_2=1-F \gtrsim 10^{-3}$ \cite{ma2023high, evered2023high, shaw2024, Tsai2025}.
We, on the other hand, obtain infidelities $E_{3,4,5} \simeq 4,7,15 \times 10^{-3}$ for $N=k+1=3,4,5$ qubit gates. 
Implementing these transformations using only two-qubit gates would require $m_{3,4,5} = 6,14,30$ CZ gates \cite{kandala2017hardware,nemkov2023efficient}. 
In the absence of error correction -- itself requiring multiqubit gates -- the resulting infidelities of the C$_{2,3,4}$Z gates would then be significantly larger, $E_{3,4,5} \gtrsim 6,14,30 \times 10^{-3}$, the more so for larger $N$. 

We have considered atoms forming star-graphs on a 2D plane, which allow the realization of $N\leq 5$ qubit gates. 
Having in sight the neutral-atom quantum computing platforms with global addressing in 3D geometries \cite{barredo2018synthetic}, it would be interesting to consider even larger graphs with extensions realizing topologically non-trivial connectivities between the qubits. 
Such structures are also relevant for solving various optimization problems that can be mapped onto the quantum Ising model realized by laser-driven atoms in reconfigurable arrays of microtraps.

\section*{Acknowledgments}

This work was supported by the EU programme HORIZON-CL4-2021-DIGITAL-EMERGING-01-30 via the Project EuRyQa (Grant No. 101070144). 
%D.P. acknowledges useful discussions with Armen Allahverdyan and support of HESC of Armenia (Grant No. 24FP-1F03).

\section*{Data availability}

The data that support the results of this article are openly
available \cite{MQgates_dist_dataset}.

\bibliographystyle{quantum}
\bibliography{refs}

\newpage

%%%%%%%%%%%%%%%%%%%%%%%%%%%%%%%%%%%%%%
%\onecolumn

\appendix

\section{Full circuits implementing the C$_k$Z gates} 

\subsection{Star-graphs}
\label{App:Correction_Circuit}

Here we show that application of X and Z gates to all but one of the qubits before the transformation $\mathcal{U}_{\mathrm{II}}\mathcal{U}_{\mathrm{I}}$, followed by X gates applied to the same qubits, as illustrated in Fig.~\ref{Fig:XZXGate_protocol}, is equivalent to the C$_k$Z gate of Eq.~(\ref{Eq:C_kZ-truth-table}) of the main text. 

Without loss of generality, we may assume that the single qubit gates are applied to all the qubits $i=1,2,\ldots,k$ except for the central qubit $i=0$. 
The combined transformation is thus 
\begin{equation} \label{Eq:XUUZX}
    \Bigg(\prod_{i=1}^k \mathrm{X}_i\Bigg)
    \mathcal{U}_{\mathrm{II}}\mathcal{U}_{\mathrm{I}}\Bigg(\prod_{i=1}^k \mathrm{Z}_i\mathrm{X}_i\Bigg)\ket{\mathbf{q}} .
\end{equation}
Proceeding one step at a time, we have 
\begin{equation} \label{Eq:ZX}
    \prod_{i=1}^k \mathrm{Z}_i\mathrm{X}_i \ket{\mathbf{q}} =  (-1)^{\sum_{i=1}^k\bar{q}_i} \ket{q_0 \bar{q}_1 \ldots \bar{q}_k} ,
\end{equation}
where $\bar{q}_i=1-q_i$. Next
\begin{equation} \label{Eq:UU}
\mathcal{U}_{\mathrm{II}}\mathcal{U}_{\mathrm{I}} \ket{q_0 \bar{q}_1 \ldots \bar{q}_k} = e^{i \nu'_{\mathbf{q}} \pi} \ket{q_0 \bar{q}_1 \ldots \bar{q}_k}
\end{equation}
where 
\begin{equation} \label{Eq:vp}
\nu'_{\mathbf{q}} =
    \begin{cases} 
        1, & \text{if }  \ket{\mathbf{q}}=\ket{11...1}\\
        \sum_{i=1}^k \bar{q}_i, & \text{otherwise}. 
    \end{cases}
\end{equation}
Finally, 
\begin{equation} \label{Eq:X}
    \prod_{i=1}^k \mathrm{X}_i  \ket{q_0 \bar{q}_1 \ldots \bar{q}_k}  =  \ket{\mathbf{q}} .
\end{equation}
Hence, the result of the combined transformation (\ref{Eq:XUUZX}) is
\begin{equation}
(-1)^{\big(\nu'_{\mathbf{q}}+\sum_{i=1}^k\bar{q}_i\big)}\ket{\mathbf{q}} = (-1)^{\prod^{k}_{i=0} q_i}\ket{q_0 q_1 \ldots q_k} ,
\end{equation}
since $\big(\nu'_{\mathbf{q}} + \sum_{i=1}^k\bar{q}_i\big) = 1$ for $\ket{\mathbf{q}}=\ket{11...1}$ and is an even number
$2 \sum_{i=1}^k\bar{q}_i$ for all the other inputs.

\subsection{Star-graphs with extended branches} \label{App:Correction_Circuit_extensions}

Here we show that the protocol of Sec.~\ref{Sec:Distant_atoms} implements C$_k$Z gates between distant atoms connected to the central atom via auxiliary atoms initially in state $\ket{1}$.  

We first derive Eq.~\eqref{Eq:nu_r-extended} for the number of Rydberg excitations $\tilde{\nu}_\mathbf{Q}$ in the MIS for a star-graph with $k$ branches, each branch $i=1,2,\ldots,k$ containing $n_i \geq 0$ auxiliary atoms.
To this end, we examine two cases: 

(i) If $q_0=0$, the central atom remains passive and the participating sub-graph consists of $k$ disconnected 1D chains (graph branches), each of length $l_i = n_i+q_i$. Then, for each branch, the MIS corresponds to an antiferromagnetic-like (AFM-like) state of $\lceil l_i/2\rceil$ Rydberg excitations at non-adjacent sites, resulting in the total number of $\tilde{\nu}_\mathbf{Q}=\sum_{i=1}^k\lceil l_i/2\rceil$ excitations. 

(ii) If $q_0=1$, we have a star-like tree graph with $k$ branches, each of length $l_i$. 
We now consider the following sub-cases:

\begin{itemize}

\item If all branches are of even length $l_i$, 
the excitation of the central atom does not change the number of excitations in each branch $\lceil l_i/2 \rceil$, and the total number of excitations is $\tilde{\nu}_\mathbf{Q}=1+\sum_{i=1}^k\lceil l_i/2\rceil$. 

\item If only one branch has odd length $l_i$, we have degenerate MIS configurations with the central atom excited or not. The latter is equivalent to case~(i), while for the former the excitation of the central atom reduces by one the number of excitations in the odd-length branch and does not affect the remaining $k-1$ even-length branches as above. 
In either case, the total number of excitations is $\tilde{\nu}_\mathbf{Q}=\sum_{i=1}^k\lceil l_i/2\rceil$.

\item If $k >1$ branches have odd length $l_i$, the central atom is not excited in the MIS state, since it would suppress the excitation of $k>1$ atoms in the odd-length branches. 
Hence, as in case (i), we again obtain total number of $\tilde{\nu}_\mathbf{Q}=\sum_{i=1}^k\lceil l_i/2\rceil$ excitations.   

\end{itemize}
Summarizing all the cases above, we have 
\begin{equation} 
\tilde{\nu}_\mathbf{Q} = 
\begin{cases}
1 + \sum_{i=1}^k\lceil l_i/2\rceil & \text{if }  q_0=1 \; \& \; l_i \text{ is even } \forall \, i \\
\sum_{i=1}^k\lceil l_i/2\rceil & \text{otherwise} 
\end{cases}
\end{equation}
which is Eq.~(\ref{Eq:nu_r-extended}) of the main text. 

Now note that when $n_i$ is odd, $l_i=n_i +q_i$ is even for $q_i=1$ and is odd for $q_i=0$, and $\lceil l_i/2\rceil = (n_i+1)/2$ for either $q_i$. 
Hence, when all $n_i$ are odd, $\tilde{n} \equiv \sum_{i=1}^k\lceil l_i/2\rceil = \frac{1}{2}(N-1)$ has the same value for all input values of $q_i$ ($i=1,2,\ldots,k$), and only the input configuration with the central qubit $q_0=1$ and all the other qubits $q_i=1$ leads to the MIS parity $(-1)^{1+\tilde{n}}$ opposite to that, $(-1)^{\tilde{n}}$, for all the other inputs. 
Thus the transformation $\mathcal{U}_{\mathrm{II}}\mathcal{U}_{\mathrm{I}}$ with the MIS intermediate state $\ket{R_{\mathbf{Q}}}$ implements the C$_k$Z gate without additional single-qubit gates. 

When some $n_i$'s are even, similarly to Sec.~\ref{App:Correction_Circuit}, we apply the X and Z gates to the qubits in the odd-$n_i$ branches before the transformation $\mathcal{U}_{\mathrm{II}}\mathcal{U}_{\mathrm{I}}$, followed by X gates on the same qubits. 
The X gates redefine the basis of the qubits in the corresponding branches, whereas the Z gates remove the trivial $\pi$-phase change. 
The combined transformation then implements the C$_k$Z gate.  

%%%%%%%%%%%%%%%%%%%%%%%%%%%%%%%%%%%%%%%%%%%%%

\section{Population and phase dynamics} \label{App:Population_phase_dynamics}

%%%%%%%%%%%%%%%%%%%%%%%%%%%%%%%%%%%
\begin{figure*}[t!]
  \centering
    \includegraphics[width=1.0\textwidth]{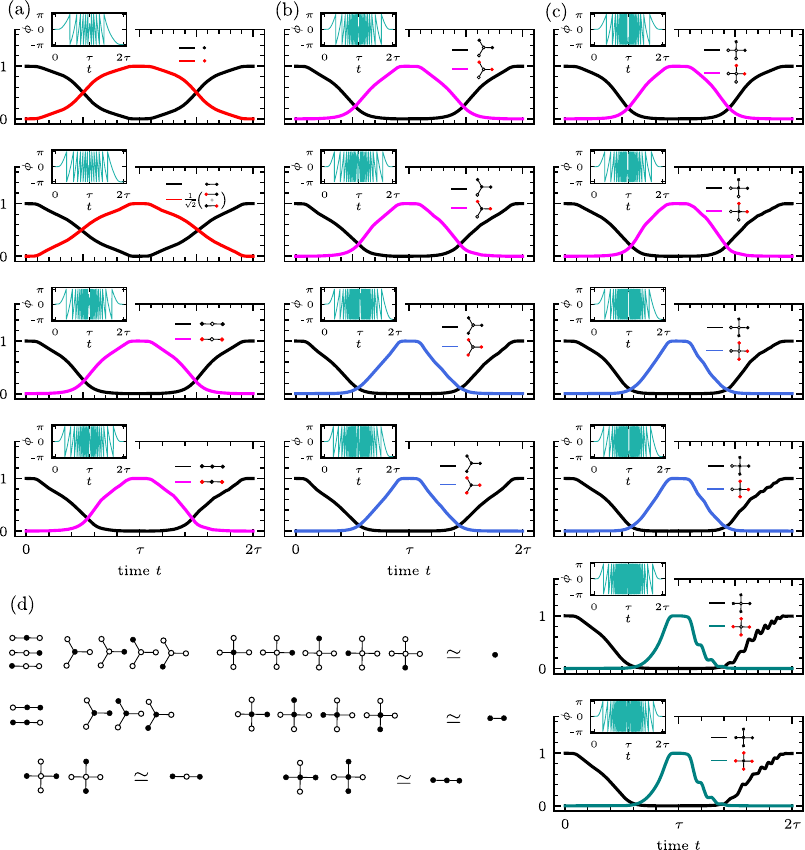} 
    \caption{Dynamics of populations of the initial input states $\ket{\mathbf{q}}$ (black lines) and the corresponding intermediate MIS states $\ket{R_\mathbf{q}}$ with $\nu_{\mathbf{q}}=1,2,3,4$ (red, magenta, blue, teal lines) Rydberg excitations during the transformation $\mathcal{U}_{\mathrm{II}}\mathcal{U}_{\mathrm{I}}$, realized by the laser pulses of Eqs.~(\ref{eq:PulseI}) and (\ref{eq:OmegaDelta}) with parameters as in Figs. \ref{Fig:Spectrum}, \ref{Fig:TrDynamics}, for implementing the C$_2$Z (a), C$_3$Z (b), and C$_4$Z (c) gates. 
    Each panel corresponds to an appropriate input configuration, and (d) illustrates equivalent configurations for different subgraphs. 
    Inset in each panel shows the dynamics of the total phase $\phi(t) = \arg \braket{\Psi(0)}{\Psi(t)}$ of the state $\ket{\Psi}$ of the system.}
    \label{Fig:Dynamics_full}
\end{figure*}
%%%%%%%%%%%%%%%%%%%%%%%%%%%%%%%%

We simulate the dynamics of the $N=k+1$ atom system using the full Hamiltonian (\ref{Eq:Hamiltonian}) realizing the transformation $\mathcal{U}_{\mathrm{II}}\mathcal{U}_{\mathrm{I}}$.
In the simulations, we take the amplitude and frequency of the chirped laser pulse during step I, $0 \leq t \leq \tau$, as 
\begin{subequations}\label{eq:PulseI}
    \begin{align}
        \Omega(t) &= \Omega_0 \frac{e^{-(t-\tau/2)^8/\sigma^8}-e^{(-\tau/2\sigma)^8}}{1-e^{-(\tau/2\sigma)^8}} \\
        \Delta(t) &= \beta \; (t-\tau/2) ,
    \end{align}  
\end{subequations}
with $\sigma=0.4 \tau$ and $\beta=2\Delta_0/\tau$, while during step II, $\tau < t \leq 2\tau$, an identical pulse satisfies the conditions (\ref{eq:OmegaDelta}), see Fig.~\ref{Fig:TrDynamics}(a). 
To insure the symmetry of the spectrum for the exact cancellation of the dynamical phases $\phi_d^{\mathrm{I}} + \phi_d^{\mathrm{II}} =0$, we assume that between the two pulses the sign of the interaction $B\to -B = B'$ is changed by transferring the atoms to an appropriate Rydberg state $\ket{r} \to \ket{r'}$, as discussed in Appendix~\ref{app:BtomB}. 

In Fig.~\ref{Fig:Dynamics_full} we show the time-dependence of populations of the input/output state $\ket{\mathbf{q}}$ and the corresponding intermediate excited MIS state $\ket{R_\mathbf{q}}$ for all the possible inputs of the system implementing the C$_k$Z gates, as well as the total phase $\phi(t) \equiv \arg \braket{\Psi(0)}{\Psi(t)} = \phi_d(t) + \phi_g$ of the state $\ket{\Psi}$ of the system.
We observe that at the end of the process, $t=2\tau$, the dynamical phase vanishes, $\phi_d(2\tau) = \phi_d^{\mathrm{I}} + \phi_d^{\mathrm{II}} = 0$, and only the geometric phase $\phi_g = \phi(2\tau) = \nu_{\mathbf{q}} \pi \mod(2\pi) = 0$ or $\pi$ remains.

%%%%%%%%%%%%%%%%%%%%%%%%%%%%%%%%%%%%%%%%%%%%%%%%

\section{``Bright'' and ``dark'' eigenstates} \label{Appendix:Dark_states}

In Fig.~\ref{Fig:Spectrum} of the main text, we show only the instantaneous  eigenstates $\ket{\alpha_n}$ of Hamiltonian (\ref{Eq:Hamiltonian}) that are ``bright'' in the sense of having non-vanishing non-adiabatic transition rates $\eta_{ln}=|\langle \alpha_l|\partial_t|\alpha_n\rangle|^2 \tau/\Delta_0 \neq 0$ from states $\ket{\alpha_{l=1,m}}$ which coincide with the input/output states $\ket{\mathbf{q}}$ for $\mp \Delta \gg \Omega$ respectively. 
Starting from the completely symmetric states, such as $\ket{\mathbf{q}}=\ket{11\ldots 1}$, we can identify the subspace of the ``dark'' eigenstates of $\mathcal{H}$ which, in the absence of decay or dephasing, are completely decoupled from the initial state and all the states reachable from it.   

\begin{figure}[t]
    \centering
    \includegraphics[width=1.0\columnwidth]{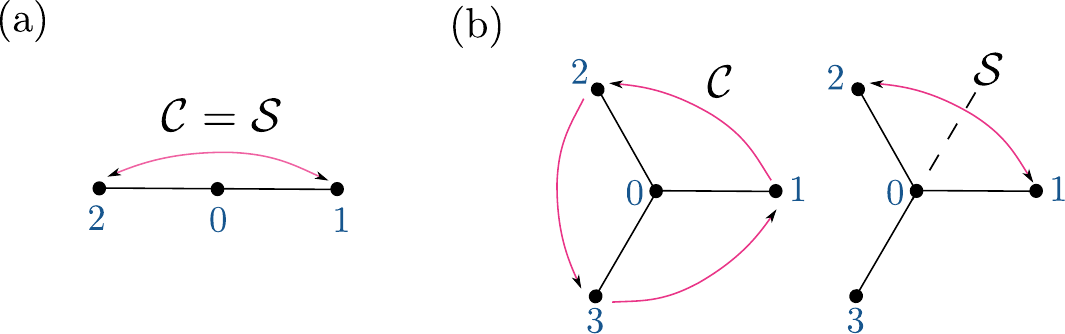}
    \caption{Symmetries of star-graphs with (a) $N=3$ and (b) $N=4$ atoms.}
    \label{Fig:SymmetriesIR}
\end{figure}

To this end, we analyze the symmetries of the  Hamiltonian \eqref{Eq:Hamiltonian}. 
We assume that the outer atoms are positioned equidistantly from the central atom, $B_{0j}= \mathrm{const} \, \forall j=1,\ldots,k$, and from their nearest neighbors, $B_{jj+1}= \mathrm{const} \, \forall \, j=1,\ldots,k$ (for $j=k$, $j+1=1$). 
The Hamiltonian is then invariant under the cyclic permutation of the outer atoms 
\begin{equation}
    \mathcal{C}=\prod^{k}_{j=1}\mathcal{S}_{jj+1}, \
\end{equation}
where 
\begin{align}
    \mathcal{S}_{ij}=&\ketbra{1_i1_j}{1_i1_j}+\ketbra{1_ir_j}{r_i1_j} \nonumber\\
    & +\ketbra{r_i1_j}{1_ir_j} + \ketbra{r_ir_j}{r_ir_j}
\end{align}
swaps atoms $i$ and $j$. 
Since $\mathcal{C}^{k}=\mathds{1}$, the eigenvalues of operator $\mathcal{C}$ are $e^{ip}$, where $p=\frac{2\pi n}{k}$ ($n=0,1...,k-1$) is the quasimomentum of the corresponding eigenstate. 
The Hamiltonian is also invariant under reflection about any axis connecting the central atom to an outer atom. 
The corresponding symmetry operators are $\mathcal{S}_{12}$ for $N=3$,  
$\mathcal{S}_{12}$ (or $\mathcal{S}_{13,23}$) for $N=4$, and $\mathcal{S}_{13}$ (or $\mathcal{S}_{24}$) for $N=5$. 
Since $\mathcal{S}^2_{ij}=\mathds{1}$, the eigenvalues of $\mathcal{S}_{ij}$ are $\lambda=\pm 1$. 
The action of operators $\mathcal{C}$ and $\mathcal{S}$ for $N=3,4$ are illustrated in Fig.~\ref{Fig:SymmetriesIR}.

%%%%%%%%%%%%%%%%%%%%%%%%%%%%%5
%%%%%%%%%%%%%%%%%%%%%%%%%%%%%%%%

\begin{figure}[t]
    \includegraphics[width=1.0\columnwidth]{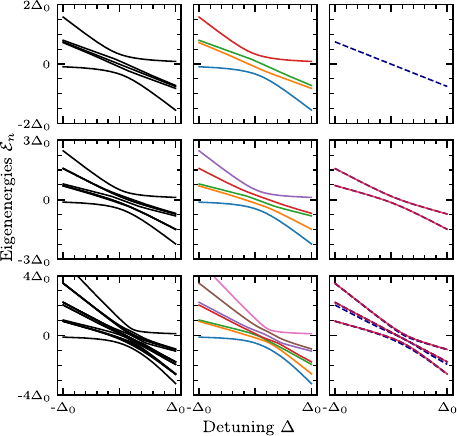}
    \caption{Adiabatic spectra of Hamiltonian vs detuning $\Delta$ for $N=3$ (upper row panels), $N=4$ (middle row panels) and $N=5$ (lower row panels). 
    For each $N$, we show the full spectrum (left panel), only the bright eigenstates with $p=0$ (middle panel), and only the dark eigenstates with $p\neq 0$ (right panel) with double degeneracy marked by an overlapping solid pink line.}
    \label{Fig:Dark_bright_345}
\end{figure}

For the symmetric initial state $\ket{\alpha_1}=\ket{\mathbf{q}}=\ket{1}^{\otimes N}$, we have $p=0$ (and $\lambda =1$). 
Since $\mathcal{C}$ commutes with the Hamiltonian $\mathcal{H}$ and its derivative $\partial_t\mathcal{H}$, only the symmetric states $\ket{\alpha_n}$ with $p=0$ can be reached from $\ket{\alpha_{l=1}}$ via 
$\bra{\alpha_l(t)} \partial_t \ket{\alpha_n(t)}=\bra{\alpha_l} \partial_t \mathcal{H}\ket{\alpha_n}/(\mathcal{E}_l -\mathcal{E}_n)$, while all the states $\ket{\alpha_{n'}}$ with $p\neq 0$ remain dark, $\bra{\alpha_l(t)} \partial_t \ket{\alpha_{n'}(t)}=0$, for any time-dependence of $\Delta(t)$ [and possibly $\Omega(t)$]. 
Hence, for $N=3$, we have one single-excitation dark state with $p=\pi$: $\ket{r^{(1)}_{\pi}} \propto \ket{1} (\ket{r1} + e^{i\pi} \ket{1r})$.
For $N=4$, the dark subspace is composed of two single-excitation states with $p=2\pi/3,4\pi/3$: 
$\ket{r^{(1)}_{p}} \propto \ket{1} (\ket{r11} + e^{i2p} \ket{1r1} + e^{ip} \ket{11r})$, and two double-excitation states with $p=2\pi/3,4\pi/3$:
$\ket{r^{(2)}_{p}} \propto \ket{1} (\ket{rr1} + e^{ip} \ket{r1r} + e^{i2p} \ket{1rr})$; out of which we can form four states: two with quasimomentum  $p=2\pi/3$, and two with quasimomentum  $p=4\pi/3$. 
Using the reflection operator $\mathcal{S}$ that commutes with the Hamiltonian and converts the state with $p=2\pi/3$ into the corresponding state with $p=4\pi/3$, and vice versa, we find that the energies of these state pairs are degenerate.    
Finally, for $N=5$, we have three single--excitation and three triple-excitation states with $p=2\pi/4,4\pi/4,6\pi/4$, and four double excitation states: one with $p=2\pi/4$, one with $p=6\pi/4$ and two with $p=4\pi/4$ (excitations at neighboring and non-neighboring atoms); out of which we can form ten states. 
Again, using the operators $\mathcal{S}$, we find three double-degenerate states, while the remaining four states are non-degenerate.  
The full spectrum involving the bright and dark eigenstates of the Hamiltonian is shown in Fig.~\ref{Fig:Dark_bright_345}.

%%%%%%%%%%%%%%%%%%%%%%%%%%%%%%%%%%%%%%%%%%%%%%
%%%%%%%%%%%%%%%%%%%%%%%%%%%%%%%%%%%%%%%%%%%%%%%

%%%%%%%%%%%%%%%%%%%%%%%%%%%%%%%%%%%%%%%%%%%%%%%

\section{Error due to leakage to wrong-parity states} \label{App:Non-adiabatic-trans}

As detailed in the main text, the geometric phase for implementing the C$_k$Z gate stems from the parity $(-1)^{\nu_{\mathbf{q}}}$ of Rydberg excitations in the intermediate state $\ket{R_{\mathbf{q}}}$. 
Non-adiabatic transitions to state(s) $\ket{B_{\mathbf{q}}}$ with the wrong parity, $\mathcal{P}\ket{B_{\mathbf{q}}}= (-1)^{\nu_{\mathbf{q}}\pm 1}\ket{B_{\mathbf{q}}}$, lead to error.
To determine the leakage error, assume that
\begin{equation}
    \mathcal{U}_{\text{I}} \ket{\mathbf{q}} 
    = a_{\mathbf{q}} \ket{R_{\mathbf{q}}}+ b_{\mathbf{q}}\ket{B_\textbf{q}},
\end{equation}
where the amplitudes $a_{\mathbf{q}}$ and $b_{\mathbf{q}}$ of the right and wrong parity states are normalized as $|a_{\mathbf{q}}|^2+|b_{\mathbf{q}}|^2=1$ and $|a_{\mathbf{q}}|^2\gg|b_{\mathbf{q}}|^2$. 
At the end of the process, at time $2\tau$, the amplitude of the initial state is then 
\begin{align}
    g_\mathbf{q} &=\braket{\mathbf{q}|\mathcal{U}_{\text{II}}\mathcal{U}_{\text{I}}}{\mathbf{q}}=\braket{\mathbf{q}|\mathcal{U}^{-1}_{\text{I}}\mathcal{P}\mathcal{U}_{\text{I}}}{\mathbf{q}}\nonumber\\
    &=(-1)^{\nu_{\mathbf{q}}}|a_{\mathbf{q}}|^2+(-1)^{\nu_{\mathbf{q}}\pm1}|b_{\mathbf{q}}|^2 \nonumber\\
    &=(-1)^{\nu_{\mathbf{q}}}\left(1-2|b_{\mathbf{q}}|^2\right),
\end{align}
and the error probability due to leakage to the wrong parity state(s) $\ket{B_{\mathbf{q}}}$ is $E_\mathbf{q}=1-|g_\mathbf{q}|^2 \simeq 4|b_{\mathbf{q}}|^2$.
Averaging over all the $2^{k+1}$ input states $\ket{\mathbf{q}}$ leads to the total leakage error 
\begin{equation}
    E_{\text{leakage}}=\frac{4}{2^{k+1}}\sum_{\mathbf{q}}|b_{\mathbf{q}}|^2 , \label{eq:app:Eleak}
\end{equation}
which is Eq.~(\ref{eq:Eleak}) of the main text. 

%%%%%%%%%%%%%%%%%%%%%%%%%%%%%%%%%%%%%%%%%%%%%%
\begin{figure}[t]
    \includegraphics[width=1.0\columnwidth]{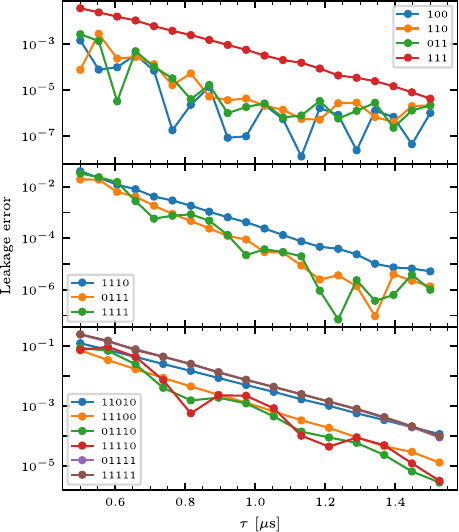}
    \caption{Leakage errors for different input states $\ket{\mathbf{q}}=\ket{q_0q_1 \ldots q_k}$ ($q_i=\{0,1\}$) for symmetric star-graphs of $N=1+k=3,4,5$ atoms (upper, middle, lower panels) vs the laser pulse duration $\tau$.}
    \label{Fig:Input_errors}
\end{figure}
%%%%%%%%%%%%%%%%%%%%%%%%%%%%%%%%%%%%%%%%%%%%%%%

In the main text, we state that during the transfer $\ket{\alpha_1}=\ket{\mathbf{q}} \to \ket{R_{\mathbf{q}}}$ the input configurations $\ket{\mathbf{q}}=\ket{q_0q_1 \ldots q_k}$ with the smallest energy gap to the nearest bright state $\ket{\alpha_2}$ with the wrong parity contribute the dominant error in Eq.~(\ref{eq:app:Eleak}). 
In Fig.~\ref{Fig:Input_errors} we show the leakage errors for various input configurations for symmetric star-graphs of $N=1+k=3,4,5$ atoms vs duration $\tau$ of the laser pulse  with the linear sweep of detuning $\Delta(t)$ as in Fig.~\ref{Fig:TrDynamics}(a1). 
For $N=3$ atoms, the dominant error is that of the input $\ket{111}$ and thus $\mu_{k=2}\simeq 1$ in Eq.~(\ref{eq:Eleak_fit}). 
For $N=4$ atoms, the three input configurations of the form $\ket{1110}$ and one of the configurations $\ket{1111}$ or $\ket{0111}$ give dominant errors, $\mu_3 \gtrsim 4$.
For $N=5$ atoms, the dominant error comes from the input configurations $\ket{11111}$ and $\ket{01111}$, 
the four configurations the form $\ket{11110}$,   
the four configurations the form $\ket{01110}$, and  
the six configurations the form $\ket{11100}$ and $\ket{11010}$, in total $\mu_4 \simeq 16$ configurations.  

%%%%%%%%%%%%%%%%%%%%%%%%%%%%%%%%%%%%%%%%%%%%%%%
%%%%%%%%%%%%%%%%%%%%%%%%%%%%%%%%%%%%%%%%%%%%%%%
\section{Changing the sign of interaction} 
\label{app:BtomB}

\begin{figure}[t] 
    \centering
    \includegraphics[width=\columnwidth]{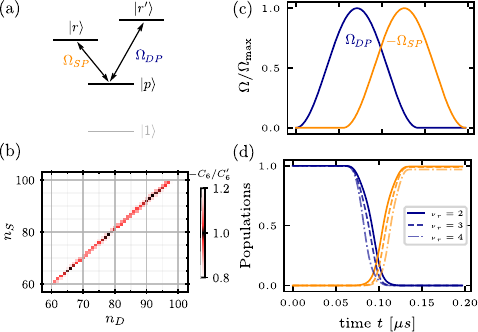} 
    \caption{(a) Level scheme for STIRAP between the Rydberg states $\ket{r}=\ket{n_S S_{1/2}}$ and $\ket{r'} = \ket{n_D D_{5/2}}$ via the intermediate state $\ket{p}=\ket{6P_{3/2}}$ of a Rb atom. 
    (b) With a proper choice of the principal quantum numbers $n_S$ and $n_D$ of states $\ket{n_S S_{1/2}}$ and $\ket{n_D D_{5/2}}$, the ratio of their vdW coefficients $-C_6/C_6'$ can be close to 1 \cite{Singer_2005}.   
    (c) Temporal profiles of the pulses 
    $\Omega_{DP}(t\leq \tau_{\mathrm{tr}}-\tau_{\mathrm{del}})=\Omega_{\max} \sin^2 [ \pi t/(\tau_{\mathrm{tr}} - \tau_{\mathrm{del}} ) ]$ and 
    $\Omega_{SP}(t\geq \tau_{\mathrm{del}})=-\Omega_{\max} \sin^2 [\pi (t-\tau_{\mathrm{del}})/(\tau_{\mathrm{tr}} - \tau_{\mathrm{del}} ) ]$ with
    $\Omega_{\max}=2\pi\times 80\,$MHz, $\tau_{\text{tr}}=0.2\,\mu$s and $\tau_{\text{del}}=0.275\tau_{\text{tr}}$. 
    (d)~Population transfer between states $\ket{R_{\mathbf{q}}}$ (blue lines) and $\ket{R'_{\mathbf{q}}}$ (orange lines) for configurations involving $\nu_{\mathbf{q}}=2$ ($\ket{R_{\mathbf{q}}}=\ket{1rr}$), $\nu_{\mathbf{q}}=3$ ($\ket{R_{\mathbf{q}}}=\ket{1rrr}$) and $\nu_{\mathbf{q}}=4$ ($\ket{R_{\mathbf{q}}}=\ket{1rrrr}$) Rydberg excitations. 
    The interaction strengths are $B_{ij} =-B'_{ij} = B/64$ for $k=2$ and $B_{ij} = B/27$ for $k=3$ with $B=2\pi \times 48\,$MHz, and $B_{ij} = B/8$ for $k=4$ with $B=2\pi \times 44.8\,$MHz, and the decay rates are $\Gamma_p=2 \pi \times 0.58\,$MHz and $\Gamma_{r,r'}=2 \pi \times 0.5\,$kHz.}
    \label{Fig:Transferrrp}
\end{figure}

We here consider in more detail the procedure for changing the sign of interaction, $B_{ij}\rightarrow-B_{ij}$, between steps I and II, which involves transferring the atoms from state $\ket{r}$ with vdW coefficient $C_6>0$ to state $\ket{r'}$ with vdW coefficient $C_6' \simeq -C_6$. 
We assume $^{87}$Rb atoms, with the ground state  $\ket{1}=\ket{5S_{1/2},F=2,M_F=0}$ coupled during step I to the Rydberg state $\ket{r}=\ket{n_S S_{1/2}}$ by a two-photon process via the intermediate state $\ket{p}=\ket{6P_{3/2}}$. 
After step I, the atoms in state $\ket{r}$ are transferred 
to the Rydberg state $\ket{r'}=\ket{n_D D_{5/2}}$ via Raman process involving two laser pulses acting on transitions $\ket{r} \to \ket{p}$ and $\ket{p} \to \ket{r'}$ with the Rabi frequencies $\Omega_{SP}$ and $\Omega_{DP}$, see Fig.~\ref{Fig:Transferrrp}(a). 
With appropriate choice of the principal quantum numbers $n_S$ and $n_D$ for the Rydberg states $\ket{r}$ and $\ket{r'}$, their vdW coefficients can be $C_6' \simeq -C_6$ (or  $C_6' = -\lambda C_6$ with $\lambda \simeq 1$), see Fig.~\ref{Fig:Transferrrp}(b).  
During step II, the atoms are returned from state $\ket{r'}$ to state $\ket{1}$ by a two-photon process via the same intermediate state $\ket{p}$.

%%%%%%%%%%%%%%%%%%%%%%%%%%%%%%%%%%%%%%%%%%
%%%%%%%%%%%%%%%%%%%%%%%%%%%%%%%%%%%%%%%%%%%%%%

Since the intermediate state $\ket{p}$ is a lower electronically excited state, it has a large decay rate $\Gamma_p \simeq 2\pi \times 0.58 \:$MHz ($\Gamma_p \gg \Gamma_{r,r'}$).  
Hence, during the Raman transfer $\ket{r} \to \ket{r'}$, we should avoid populating the intermediate state $\ket{p}$ which would result in population loss. 
This can be achieved either by using largely detuned, $|\Delta_P| \gg \max[ \Omega_{SP,DP}]$, temporally overlapping laser pulses \cite{Doultsinos2025PRR}, 
or by using resonant pulses and employing stimulated Raman adiabatic passage (STIRAP) in which the pulses $\Omega_{SP}(t)$ and $\Omega_{DP}(t)$ are applied in a ``counter-intuitive'' order \cite{RevModPhys.70.1003,RevModPhys.89.015006}. 
We here explore the latter approach. 

Let us first recall the STIRAP for a single non-interacting atom. 
Assuming resonant pulses $\Omega_{SP,DP}$, the Hamiltonian for an atom, in the frame rotating with the frequencies $\omega_{rp,r'p}$, is ($\hbar=1$)
\begin{equation}
\mathcal{H}_{\mathrm{at}}= \tfrac{1}{2}\Omega_{SP}(t)\ketbra{r}{p} + \tfrac{1}{2}\Omega_{DP}(t)\ketbra{r'}{p} + \mathrm{H.c.}
\end{equation}
The eigenstates of this Hamiltonian are 
$\ket{\psi_0}=\cos\theta\ket{r} - \sin\theta\ket{r'}$ and $\ket{\psi_{\pm}}=\frac{1}{\sqrt{2}}(\sin\theta\ket{r}\pm\ket{p}+\cos\theta\ket{r'})$ with the mixing angle $\theta$ defined via $\tan\theta = \Omega_{SP}/\Omega_{DP}$, and the corresponding energy eigenvalues are $\varepsilon_{0}= 0$ and $\varepsilon_{\pm} = \pm \tfrac{1}{2} \sqrt{|\Omega_{SP}|^2+|\Omega_{DP}|^2}$. 
The zero energy ``dark'' eigenstate $\ket{\psi_0}$ does not contain the rapidly decaying state $\ket{p}$, while the ``bright'' eigenstates $\ket{\psi_{\pm}}$ do contain $\ket{p}$ and thus are unstable. 
Our aim is to completely transfer the population 
between the two long-lived ($\Gamma_{r,r'} \ll \Gamma_p$) Rydberg states $\ket{r}$ and $\ket{r'}$ without  
populating the unstable state $\ket{p}$, which is achieved by
adiabatically changing the dark state superposition.
With the system initially in state $\ket{r}$, we first turn on the $\Omega_{DP}(t)$ field, resulting in $\braket{r}{\psi_0} = 1$ ($\Omega_{SP}(t) \ll \Omega_{DP}(t)$ and therefore $\theta(t) = 0$).
This is then followed by switching on $\Omega_{SP}(t)$ and 
switching off $\Omega_{DP}(t)$ [see Fig.~\ref{Fig:Transferrrp}(c)], 
resulting in $|\braket{r'}{\psi_0}| = 1$
($\Omega_{SP}(t) \gg \Omega_{DP}(t)$ and therefore $\theta(t) = \pi/2$). 
If the mixing angle is rotated slowly enough, 
$\dot{\theta} \ll |\varepsilon_{\pm} - \varepsilon_0|$, the system adiabatically follows the dark state $\ket{\psi_0}$, and the bright states $\ket{\psi_{\pm}}$, and thereby $\ket{p}$, are never populated. 
Hence, the decay of $\ket{p}$ is neutralized and 
the population of the system is completely transferred from $\ket{r}$ to $\ket{r'}$ [Fig.~\ref{Fig:Transferrrp}(d)].
Note that since during the transfer we follow the eigenstate with energy $\varepsilon_0=0$, the dynamical phase vanishes, and we can choose the sign of the Rabi frequency $\Omega_{SP}$ to be opposite to that of $\Omega_{DP}$ ($\pi$ phase difference) to transfer $\ket{r} \to \ket{r'}$ without sign change.  

Consider now the multiatom system initially in state $\ket{R_{\mathbf{q}}}$ with $\nu_{\mathbf{q}}$ Rydberg excitations and our aim is to transfer them to state $\ket{R'_{\mathbf{q}}}$ with all $\ket{r_i} \to \ket{r'_i}$. 
The interaction between the Rydberg state atoms is described by 
\begin{eqnarray}
\mathcal{H}_{\text{int}} &=& \sum_{i<j} \big[B_{ij}\ketbra{r_ir_j}{r_ir_j} + B'_{ij}\ketbra{r_i'r_j'}{r_i'r_j'}
\nonumber \\ & & \qquad
\tilde{B}_{ij} (\ketbra{r_i'r_j}{r_i'r_j} + \ketbra{r_i r_j'}{r_i r_j'} ) \big] , \quad
\end{eqnarray}
where $B'_{ij} = -\lambda B_{ij}$ with $\lambda \simeq 1$, and $\tilde{B}_{ij} = \chi B_{ij}$ with $|\chi| \lesssim 1$.
The total Hamiltonian is 
\begin{equation}
\mathcal{H} = \sum_i \mathcal{H}_{\mathrm{at}}^{(i)} + \mathcal{H}_{\text{int}} ,
\end{equation}
and we expect that during the STIRAP each atom will still adiabatically follow the corresponding dark state $\ket{\psi_0^{(i)}}$, provided the interactions are small compared to the energy separation from the bright states, $|\varepsilon_{\pm}| \simeq \tfrac{1}{2} \max |\Omega_{SP,DP}| \gg B_{ij} \; \forall \; i,j \geq 1$.   
We verified these conclusions by performing exact numerical simulations of the dynamics of the system evolving under the effective non-Hermitian Hamiltonian 
\[
\tilde{\mathcal{H}}= \mathcal{H} -\frac{i}{2} \left[\Gamma_r\ketbra{r_i}{r_i} + \Gamma_{r'}\ketbra{r'_i}{r'_i} + \Gamma_p\ketbra{p_i}{p_i}\right]
\]
that includes the decay of all the atomic states.  
Using strong Raman pulses $\Omega_{SP}(t)$ and $\Omega_{DP}(t)$ of proper duration $\tau_{\mathrm{tr}} - \tau_{\mathrm{del}}$ and delay $\tau_{\mathrm{del}}$ with respect to each other, we obtain nearly complete transfer between states $\ket{R_{\mathbf{q}}}$ and $\ket{R'_{\mathbf{q}}}$ involving $\nu_{\mathbf{q}} \leq 2,3,4$ Rydberg excitations for $k=2,3,4$ graphs, see Fig.~\ref{Fig:Transferrrp}(c,d). 

The transfer error for the C$_k$Z gate averaged over all the input states $\ket{\mathbf{q}}$ is 
\begin{equation}
E_{\mathrm{tr}}= \frac{1}{2^{k+1}} \sum_{\mathbf{q}} | \bra{R'_{\mathbf{q}}}\mathcal{U}_{\mathrm{tr}} \ket{R_{\mathbf{q}}} |^2,
\end{equation}
where $\mathcal{U}_{\mathrm{tr}}= \exp \left(-i\int_0^{\tau_{\mathrm{tr}}} \tilde{\mathcal{H}} dt \right)$. We vary the transfer time $\tau_{\mathrm{tr}}$ and pulse delay $\tau_{\mathrm{del}}$ to determine their optimal values that minimize the transfer error $E_{\mathrm{tr}}$, as shown in Fig.~\ref{Fig:Fidelity_345}(b) of the main text for C$_2$Z gate. 
There we assumed $\lambda =1$ and set for simplicity $\chi=0$, which, for $\Omega_{SP}(\tau_{\mathrm{tr}}-t)=-\Omega_{DP}(t)$, leads to vanishing dynamical phase.   

More generally, assuming that during the transfer each Rydberg atom $i=1,2,\ldots,\nu_{\mathbf{q}}$ adiabatically follows the state $\ket{\phi_0^{(i)}}$, the instantaneous interaction energy of the many-atom dark state $\ket{\Psi_0}=\prod_{i=1}^{\nu_{\mathbf{q}}}\ket{\psi^{(i)}_0}$ is
\begin{align}
 \mathcal{E}_{\mathrm{int}}(t) &= \bra{\Psi_0}\mathcal{H}_{\text{int}}\ket{\Psi_0} \nonumber \\
  &= \sum^{\nu_{\mathbf{q}}}_{i<j}B_{ij} %\nonumber \\   &
 \left[ \cos^4\theta(t)-\lambda\sin^4\theta(t) \right. \nonumber \\
  & \left.  \qquad \qquad +2 \chi\cos^2\theta(t)\sin^2\theta(t) \right] ,
\end{align}
leading to the dynamical phase 
$\phi_{\mathrm{int}}=\int_0^{\tau_{\mathrm{tr}}} \mathcal{E}_{\mathrm{int}}(t) dt$. 
By appropriately shaping the temporal profiles of the pulses $\Omega_{SP}(t)$ and $\Omega_{DP}(t)$, and thereby $\theta(t)$, such that 
\begin{align} 
%\int_0^{\tau_{\mathrm{tr}}} \!\! \left[ \cos^2\theta(t)-\chi\sin^2\theta(t) \right]^2 dt
& \int_0^{\tau_{\mathrm{tr}}} \!\! \left[ (1 - \chi) \cos^2\theta(t)+\chi \right]^2 dt \nonumber \\
& = \int_0^{\tau_{\mathrm{tr}}} \!\! (\chi^2+\lambda ) \sin^4 \theta(t) dt ,
\end{align}
we can make the dynamical phase vanish, $\phi_{\mathrm{int}} =0$.
Obviously, for $\chi=0$ and $\lambda =1$, the above condition reduces to $\int_0^{\tau_{\mathrm{tr}}} \cos^4 \theta(t) dt = \int_0^{\tau_{\mathrm{tr}}} \sin^4 \theta(t) dt$ satisfied by the pulses $\Omega_{SP}(\tau_{\mathrm{tr}}-t)=-\Omega_{DP}(t)$ as mentioned above.

\end{document}